\documentclass[a4paper]{article}
\usepackage{hyperref} 
\usepackage{lineno}
\usepackage{authblk}
\usepackage[T1]{fontenc}
\usepackage[hmargin=2.5cm]{geometry}
\usepackage{apacite}
\usepackage[round,sort,comma,numbers]{natbib}
\setcitestyle{authoryear}

\usepackage{amssymb,amsmath}

\DeclareMathOperator{\EX}{\mathbb{E}}
\usepackage{ctable} 

\AtBeginDocument{}

\begin{document}

\title{HMCLab: a framework for solving diverse geophysical inverse problems using the Hamiltonian Monte Carlo method}

\author[1]{Andrea Zunino$^{\dagger}${\footnote{Corresponding author: Andrea Zunino \href{mailto:andrea.zunino@erdw.ethz.ch}{andrea.zunino@erdw.ethz.ch}}} }
\author[1]{Lars Gebraad$^{\dagger}$}
\author[2]{Alessandro Ghirotto}
\author[1]{Andreas Fichtner}
\affil[1]{\small Department of Earth Sciences, ETH Zurich, Switzerland}
\affil[2]{\small Applied Geophysics Laboratory, DISTAV University of Genoa, Italy}

\date{\today}



\maketitle


\begin{abstract}
  The use of the probabilistic approach to solve inverse problems is becoming more popular in the geophysical community, thanks to its ability to address nonlinear forward problems and to provide uncertainty quantification. However, such strategy is often tailored to specific applications and therefore there is a lack of a common platform for solving a range of different geophysical inverse problems and showing potential and pitfalls of the methodology.
  In this work, we demonstrate a common framework within which it is possible to solve such inverse problems ranging from, e.g, earthquake source location to potential field data inversion and seismic tomography. 
  This approach, in fact, can provide probabilities related to certain properties or structure of the subsurface, such as histograms of the value of some physical property, the expected volume of buried geological bodies or the probability of having boundaries defining different layers.
    Thanks to its ability to address high-dimensional problems, the Hamiltonian Monte Carlo (HMC) algorithm has emerged as the state-of-the-art tool for solving geophysical inverse problems within the probabilistic framework. HMC requires the computation of gradients, which can be obtained by adjoint methods. This unique combination of HMC and adjoint methods is what makes the solution of tomographic problems ultimately feasible.
  These results can be obtained with ``HMCLab'', a numerical laboratory for solving a range of different geophysical inverse problems using sampling methods, focusing in particular on the HMC algorithm. HMCLab consists of a set of samplers (HMC and others) and a set of geophysical forward problems. For each problem its misfit function and gradient computation are provided and, in addition, a set of prior models can be combined to inject additional information into the inverse problem. This allows users to experiment with probabilistic inverse problems and also address real-world studies. We show how to solve a selected set of problems within this framework using variants of the HMC algorithm and analyze the results. HMCLab is provided as an open source package written both in Python and Julia, welcoming contributions from the community.
\end{abstract}


\section{Introduction}

In the probabilistic approach, an inverse problem essentially represents an indirect measurement where the knowledge about the observed data and model parameters is completely expressed in terms of probabilities \citep{tarantolaInverseProblemTheory2005a}. Within such formalism, the general solution to the inverse problem is a probability density function (PDF), i.e., the posterior PDF (see \citet{tarantolaInverseProblemsQuest1982a} and \citet{mosegaardProbabilisticApproachInverse2002} for a detailed explanation).
The posterior PDF is constructed from the combination of two pieces of separate information: 1) the prior knowledge on the model parameters, expressed by the PDF $\rho(\mathbf{m})$, where $\mathbf{m}$ represents the model parameters and 2) the information provided by the experiment, described by $L(\mathbf{m})$. The posterior distribution, under certain fairly wide assumptions, is then given by \citep{mosegaardProbabilisticApproachInverse2002,tarantolaInverseProblemTheory2005a}:
\begin{linenomath*}
\begin{align}
  \sigma(\mathbf{m}) = k \, \rho(\mathbf{m}) \, L(\mathbf{m}).
\end{align}
\end{linenomath*}
Since $\sigma(\mathbf{m})$ is a PDF, it requires to evaluate the relevant integrals to find features of interest. For example, calculating the expected model given the data requires evaluating the following integral
\begin{linenomath*}
\begin{equation}
  \label{eq:mcmcexpectation}
  \EX \left[ \mathbf{m} \right] = \int_{M} \mathbf{m} \, \sigma(\mathbf{m}) \, \mathrm{d}\mathbf{m},
\end{equation}
\end{linenomath*}
where $M$ represents the whole model space.

In the particular case of linear forward models and Gaussian uncertainty, the probabilistic formalism provides the same closed-form solution than the classical least squares approach. In general, however, such high-dimensional integrals cannot be computed. Therefore, we resort to \emph{sampling}, a technique to approximate the computation of high-dimensional integrals. By generating points in the model space whose density (number of points per unit volume) is proportional to the posterior $\sigma(\mathbf{m})$, i.e., samples, one greatly reduces the amount of computations required to estimate statistics (such as $\EX \left[ \mathbf{m} \right]$) compared to a systematic grid search.
Markov Chain Monte Carlo (MCMC) methods provide a clever way to construct a Markov chain that produces samples drawn for the target distribution, i.e., the posterior PDF.
Statistical analysis of the samples obtained with MCMC then provides the answer to any inquiry in terms of probability of certain events, i.e., specific features of the solution. Practically, this means we can compute probabilities related to particular properties or structures of the solution. For instance, we might be interested in the probability of a certain geological body to have a certain volume, or the probability that there is a continuous permeable layer connecting two locations in the subsurface. \\
MCMC to sample target distributions has evolved from the appearance of the original Metropolis algorithm in physics \cite[e.g.,][]{metropolisMonteCarloMethod1949,metropolisEquationStateCalculations1953a} into a plethora of variants in many different scientific fields, including geophysics \cite[see ][for a review]{sambridgeMonteCarloMethods2002}. The MCMC method to solve inverse problems in geophysics has a long history, which dates back to the pioneering work of \citet{keilis-borokInverseProblemsSeismology1967} and \citet{pressEarthModelsObtained1968}, which were contemporary to the classic work of \citet{backusNumericalApplicationsFormalism1967a} on geophysical inverse theory. The formalization of a comprehensive theoretical framework based on probability theory for geophysical problems was started in the 80s by \citet{tarantolaInverseProblemsQuest1982a} and extended later on to include MCMC sampling methods to solve nonlinear inverse problems  \cite[e.g.,][]{mosegaardMonteCarloSampling1995,mosegaardProbabilisticApproachInverse2002,tarantolaInverseProblemTheory2005a,mosegaardQuestConsistencySymmetry2011}. An extensive review can be found in \citet{debskiChapterProbabilisticInverse2010}.

There has been a recent increase in the use of Monte Carlo methods to solve (geophysical) inverse problems, essentially for two reasons: I) relatively recent advances in the sampling algorithms and II) substantial increase in the available computational resources. These advances now allow us to tackle medium (hundreds of model parameters) to relatively large-scale (tens of thousands model parameters) inverse problems within the probabilistic framework. 
Regarding I), the literature provides a large collection of generic algorithms to perform Monte Carlo sampling, including classic MCMC \cite[e.g.,][]{hastingsMonteCarloSampling1970}, slice sampling \cite[e.g.,][]{nealSliceSampling2003}, Gibbs sampling \cite[e.g.,][]{gemanStochasticRelaxationGibbs1984},  rejection sampling \cite[e.g.,][]{gilksAdaptiveRejectionMetropolis1995}, sequential Monte Carlo \cite[e.g.,][]{liuSequentialMonteCarlo1998} and trans-dimensional Monte Carlo \cite[e.g.,][]{greenReversibleJumpMarkov1995a}, just to cite some of the main categories. The amount of literature dedicated to such algorithms from different fields is so vast that it would be pointless to attempt to give an overview here. More specifically, in geophysics there has been a constant increase in the number of algorithms proposed both for (pseudo-)sampling the posterior PDF and performing global optimization. These include, for instance, an extension of the classic Metropolis-Hasting sampler \cite[e.g.,][]{mosegaardMonteCarloSampling1995},
simulating annealing for global optimization \cite[e.g.,][]{stoffaNonlinearMultiparameterOptimization1991},
a strategy based on the nearest neighbour (Voronoi) partition of the model space  \cite[e.g.,][]{sambridgeGeophysicalInversionNeighbourhood1999},
joint inversion of different geophysical data with a cascade MCMC \cite[e.g.,][]{boschLithologicTomographyPlural1999},
trans-dimensional MCMC inversion \cite[e.g.,][]{malinvernoParsimoniousBayesianMarkov2002,bodinSeismicTomographyReversible2009} and samplers with geostatistical-based priors \cite[e.g.,][]{hansenInverseProblemsNontrivial2012a,zuninoMonteCarloReservoir2015}.
Regarding II), the geophysicist's arsenal for solving inverse problems now includes large high-performance computing resources and more easily accessible computation accelerators such as GPUs, TPUs and high-thread-count CPUs.

Traditional algorithms such as the random walk Metropolis may find difficult to setup an efficient sampler for large problems because of two reasons. The first is the property of generating correlated samples which requires a large number of iterations to obtain reliable statistics. The second is the difficulty for the proposal mechanism to generate high-probability models. In high-dimensional spaces, simply randomly perturbing a model will be very unlikely to produce another model with a higher posterior PDF. The reason is that high-dimensional spaces tend to be very empty (``curse of dimensionality'') and so the vast majority of search directions (random perturbations) will point to low probability regions, making the overall algorithm inefficient \citep{curtisPriorInformationSampling2001a}. 
If strong prior information is available and if sampling directly the prior is a possibility, the extended Metropolis algorithm \citep{mosegaardMonteCarloSampling1995} can offer a substantial improvement in terms of efficiency. Nevertheless, defining a geologically realistic prior and being able to sample it may not be an easy task in practice. 
Other approaches are based on an adaptive algorithm which, for instance, computes an approximate local covariance matrix and then samples such information to increase the chance of moving in a direction of higher probability \cite[e.g.,][]{gilksMarkovChainMonte1996}.
A recently proposed methodology for improving the proposal strategy is that of constructing an ad hoc proposal PDF based on the results of a simplified deterministic inversion \citep{khoshkholghInformedProposalMonte2021,khoshkholghFullwaveformInversionInformedproposal2022}, which will make the sampling more efficient in practice, without altering the final equilibrium distribution. This may enable the solution of large problems, although such methodology requires solving an additional inverse problem beforehand and performing some sort of interpretation in addition to the estimation of the modeling error.

The Hamiltonian Monte Carlo (HMC) method has recently gained attention in solid Earth geophysics because of its peculiar properties \citep{duaneHybridMonteCarlo1987,nealMCMCUsingHamiltonian2012,fichtnerHamiltonianMonteCarlo2019}. It combines sampling with ideas from the field of optimization, where the proposal mechanism exploits also information coming from the gradient of the posterior distribution. This unique combination enables a more efficient solution of problems when the calculation of gradients is not computationally too expensive with respect to the cost of simulating the forward problem, e.g., when adjoint methods \citep{tarantolaInversionSeismicReflection1984,trompSeismicTomographyAdjoint2005a,fichtnerAdjointMethodSeismology2006a,plessixReviewAdjointstateMethod2006} come into play \cite[e.g.,][]{zuninoIntegratingGradientInformation2018,fichtnerHamiltonianMonteCarlo2019,gebraadBayesianElasticFullWaveform2020}. This is effectively a result of the No Free Lunch-theorem described by \citet{wolpertNoFreeLunch1997}: one applies prior knowledge to the objective function and its properties (i.e., cheaply available gradient information), whereby it becomes possible to select a relatively efficient algorithm.
HMC is capable of generating more uncorrelated samples compared to traditional purely random-walk based algorithms such as the random walk Metropolis algorithm \citep{nealMCMCUsingHamiltonian2012}, producing more accurate statistical estimations with a smaller number of sampler. Thanks to this property, HMC is more suitable to address high-dimensional inverse problems than traditional derivative-free sampling methods. In fact, the cost of generating independent samples with HMC under increasing dimension $n$ grows as $O(n^{5/4})$ \citep{nealMCMCUsingHamiltonian2012}, whereas it grows as $O(n^2)$ for standard Metropolis-Hastings \citep{creutzGlobalMonteCarlo1988}

An alternative and recently popularized approach using also information from the gradient is Stein Variational Gradient Descent \cite[SVGD,]{liuSteinVariationalGradient2019} a variational inference algorithm, which aims at approximate inference by minimizing the Kullback-Leibler divergence between the proposed and target distributions.

Although the theoretical formalism and the infrastructure to perform intensive computations are there, a common framework to address different geophysical inverse problems has not emerged yet. Implementations of the HMC algorithm are typically application-specific and often not easily accessible to non-specialists. In addition, as these methods are nascent in the field of solid Earth geophysics, the community as a whole has not had time to acquire substantial expertise in the usage of these methods in order to evaluate their potential and routinely apply them to realistic problems.  
This work aims at facilitating at least part of the generation of this expertise, specifically in applying gradient-based sampling methods to inverse problems and analyzing their results. In this work we show how HMC can be used to obtain useful information from a set of diverse geophysical data sets through some illustrative selected examples from seismology and potential fields problems. All problems are addressed within the same framework, where generic samplers and data structures allows us to easily experiment with different data, priors and possibly to combine them.

These results are obtained with  ``HMCLab'', a tool to solve research problems and a numerical laboratory for experimenting with inverse algorithms such as HMC for a variety of geophysical topics. HMCLab provides software for a set of geophysical problems, for which functions to solve the forward problem, compute gradients of the misfit function and  several kinds of priors and samplers for the HMC method are provided. This package is currently written partly in Python~\citep{vanrossumPythonTutorial1995} and partly in Julia~\citep{bezansonJuliaFreshApproach2017}, depending on the specific problem. It is, however, in constant evolution as new geophysical problems are added or translated into the other one of the two languages. Moreover, users can supply their own forward model functions and priors which can easily be used with the HMC samplers. In addition, several Jupyter Notebooks are provided that guide the user through the various aspects of applying MCMC algorithms and analyzing their results, for various inverse problems. \\
In the following, we first give a brief overview of the core of HMC algorithms and illustrate what kind of information can obtained by solving some selected example problems using HMCLab.

\section{Theoretical background}

\subsection{The original HMC sampler}
\label{sec:hmcalgo}

HMC constructs a Markov chain over an $n$-dimensional probability density function $\sigma(\mathbf{m})$ using classical Hamiltonian mechanics. The algorithm regards the current state $\mathbf{m}$ of the Markov chain as the location of a physical particle in an $n$-dimensional space $\mathcal{M}$ (i.e., model or parameter space). It moves under the influence of a potential energy, $U$, which is defined as
\begin{linenomath*}
    \begin{align}
        U(\mathbf{m})=-\ln{ \left( \sigma(\mathbf{m}) \right) }.
    \end{align}
\end{linenomath*}
To complete the physical system, the state of the Markov chain needs to be artificially augmented with momentum variables $\mathbf{p}$ and a generalized mass for every dimension pair. The collection of resulting masses is contained in a symmetric positive definite mass matrix $\mathbf{M}$ of dimension $n \times n$. The momenta and the mass matrix define the kinetic energy of the particle as
\begin{linenomath*}
\begin{align}
    K(\mathbf{p})=\frac{1}{2} \mathbf{p}^T \mathbf{M}^{-1} \mathbf{p}.
\end{align}
\end{linenomath*}
In the HMC algorithm, the momenta $\mathbf{p}$ are drawn randomly from a multivariate Gaussian with covariance matrix $\mathbf{M}$ (the mass matrix). The sum of  the location-dependent potential and momentum-dependent kinetic energy constitute the total energy, or Hamiltonian, of the system
\begin{linenomath*}
\begin{align}
    H(\mathbf{m},\mathbf{p})=U(\mathbf{m})+K(\mathbf{p}).
\end{align}
\end{linenomath*}
The Hamiltonian dynamics are governed by the following equations,
\begin{linenomath*}
\begin{align}
    \frac{\partial\mathbf{m}}{\partial\tau} = \frac{\partial H}{\partial\mathbf{p}},\quad \frac{\partial\mathbf{p}}{\partial\tau} = - \frac{\partial H}{\partial\mathbf{m}} \, ,
\end{align}
\end{linenomath*}
which determine the position and momentum of the particle as a function of time $\tau$. This time $\tau$ is artificial just like the mass matrix, it has no connection to the actual physics of the inverse problem at hand.

We can simplify Hamilton's equations using the fact that kinetic and potential energy depend only on momentum and location, respectively, to obtain
\begin{linenomath*}
\begin{align}
    \frac{\partial\mathbf{m}}{\partial\tau} = \mathbf{M}^{-1} \mathbf{p}, \quad \frac{\partial\mathbf{p}}{\partial\tau} = - \frac{\partial U}{\partial\mathbf{m}} \, .
\end{align}
\end{linenomath*}
Evolving $\mathbf{m}$ over time $\tau$ generates another possible state of the system with new position $\mathbf{\tilde{m}}$, momentum $\mathbf{\tilde{p}}$, potential energy $\tilde{U}$, and kinetic energy $\tilde{K}$. Due to the conservation of energy, the Hamiltonian is equal in both states, i.e., $U+K = \tilde{U} + \tilde{K}$. Successively drawing random momenta and evolving the system generates a distribution of the possible states of the system. Thereby, HMC samples the joint momentum and model space, referred to as phase space. As we are not interested in the momentum component of phase space, we marginalize over the momenta by simply dropping them. This results in samples drawn from $\sigma(\mathbf{m})$.

If one could solve Hamilton's equations exactly, every proposed state (after burn-in) would be a valid sample of $\sigma(\mathbf{m})$. Since Hamilton's equations for non-linear forward models cannot be solved analytically, the system must be integrated numerically. Suitable integrators are symplectic, meaning that time reversibility, phase space partitioning and volume preservation are satisfied \citep{nealMCMCUsingHamiltonian2012,fichtnerHamiltonianMonteCarlo2019}. In this work, we employ the leapfrog method as described in \citet{nealMCMCUsingHamiltonian2012}, with higher order symplectic integrators also implemented. However, the Hamiltonian is generally not preserved exactly when explicit time-stepping schemes are used \cite[e.g.,][]{simoExactEnergymomentumConserving1992c}. Therefore, the time evolution generates samples not exactly proportional to the original distribution. A Metropolis-Hastings correction step is therefore applied at the end of numerical integration.

In summary, at each iteration, samples are generated starting from a randomly drawn model $\mathbf{m}$ in the following way:
\begin{enumerate}
   \item Propose momenta $\mathbf{p}$ according to the Gaussian with mean $\mathbf{0}$ and covariance matrix $\mathbf{M}$;
   \item Compute the Hamiltonian $H$ of model $\mathbf{m}$ with momenta $\mathbf{p}$;
   \item Propagate $\mathbf{m}$ and $\mathbf{p}$ for some time $\tau$ to $\tilde{\mathbf{m}}$ and $\tilde{\mathbf{p}}$, using the discretized version of Hamilton's equations and a suitable numerical integrator;
   \item Compute the Hamiltonian $\tilde{H}$ of model $\tilde{\mathbf{m}}$ with momenta $\mathbf{\tilde{p}}$;
   \item Accept the proposed move $\mathbf{m} \rightarrow \tilde{\mathbf{m}}$ with probability
      \begin{eqnarray}
          p_\text{accept} = \min \left( 1, \exp ( H-\tilde{H} ) \right)\,.
      \end{eqnarray}
   \item If accepted, use (and count) $\tilde{\mathbf{m}}$ as the new state. Otherwise, keep (and count) the previous state. Then return to 1.
\end{enumerate}

The mass matrix $\mathbf{M}$ is one of the important tuning parameters of the HMC algorithm; details on its meaning and suggestions for tuning can be found in \citet{fichtnerHamiltonianMonteCarlo2019,fichtnerAutotuningHamiltonianMonte2021}. Moreover, employing the discrete leapfrog integrator implies that there are two additional parameters that need to be tuned, namely the time step $\epsilon$
and the number of iterations $L$ \citep{nealMCMCUsingHamiltonian2012}.

\subsection{HMC variants}

The algorithm described so far is the simplest version of HMC, however, several variants of the original algorithm exist, which mostly aim at automatically tuning some of the parameters or improving mixing \citep{nealMCMCUsingHamiltonian2012,sambridgeParallelTemperingAlgorithm2014,fichtnerAutotuningHamiltonianMonte2021}.

A notable example is the No U-Turn Sampler (NUTS) \citep{hoffmanNoUturnSamplerAdaptively2014}, which aims at providing an automatic tuning of the two leapfrog integrator-related parameters, $\epsilon$ and $L$. NUTS finds a suitable value for $\epsilon$ during the burn-in and then fixes it for the following iterations to avoid breaking the detailed balance property. The number of iterations $L$ instead is dynamically adjusted (``dynamic HMC'') at each iteration in a way such that there is no doubling back of the trajectory. This allows for long or short moves depending on the region of the model space which the algorithm is visiting. NUTS is implemented in HMCLab following \citet{hoffmanNoUturnSamplerAdaptively2014}.

Additionally, methods to investigate inverse problems that might show strongly isolated modes exist. By running multiple chains with tempered (i.e. smoothed) posterior PDFs and letting these samplers exchange states, the exploration of local minima might be accelerated \citep{sambridgeParallelTemperingAlgorithm2014}. Tempered trajectories following \citet{nealMCMCUsingHamiltonian2012} may also help discovering isolated modes. This variation does not require multiple Markov chains, nonetheless, it is able to more easily transition between local minima, at the expense of a reduced acceptance rate.

\subsection{Gradient computations}

As mentioned above, one important aspect of a successful HMC strategy is the capability to efficiently compute the gradient of the potential energy $\nabla U(\mathbf{m}) = \nabla \! \left( -\log(\sigma(\mathbf{m})) \right)$. The first method that proves powerful for relatively simple models is to evaluate derivatives analytically. This typically allows for cheap computation of the gradients, but is only applicable to models that can be analytically differentiated. This is most notably used for the joint non-linear source location and medium velocity estimation as mentioned later in this manuscript. For larger problems, a tool that can provide a substantial help in making gradient calculations efficient is the adjoint method \cite[e.g.,][]{lionsOptimalControlSystems1971,tarantolaInversionSeismicReflection1984,talagrandVariationalAssimilationMeteorological1987,trompSeismicTomographyAdjoint2005a,fichtnerAdjointMethodSeismology2006a,plessixReviewAdjointstateMethod2006,hinzeOptimizationPDEConstraints2008}. This strategy allows us to compute the gradient $\nabla U$ with a computational cost of about two (three in practice) forward simulations, much cheaper than other approaches such as finite difference methods. We employ the adjoint technique in some of our geophysical problems, namely in the case of acoustic and elastic full waveform inversion and for the nonlinear traveltime problem (eikonal solver). 

Another useful tool to compute the gradient for certain problems is automatic  differentiation \cite[e.g.,][]{sambridgeAutomaticDifferentiationGeophysical2007,griewankEvaluatingDerivativesPrinciples2008}, a computational technique where derivatives of a user-coded function are provided automatically by the software in the form of a function. This technique can be convenient for problems where it is difficult to derive the adjoint equations (e.g., when the forward operator is not self-adjoint) or where the forward model needs to be adapted for each specific case because it depends, e.g., on the specific rock types present in the area under study, requiring a re-derivation of the analytical derivatives (such as rock physics models \citep{mavkoRockPhysicsHandbook2003a}). We use this tool, e.g., for the problem of inversion of amplitude-versus-angle (AVA) seismic reflection data, where the forward modelling is a combination of a rock physics model \cite[e.g.,][]{mavkoRockPhysicsHandbook2003a} and a convolutional seismic model. 

\subsection{Prior information}

Prior information plays an important role in solving inverse problems by providing additional information directly on the model parameters to better constrain plausible values for the solution and helping to mitigate the non-uniqueness \cite[e.g.,][]{curtisPriorInformationSampling2001,scalesPriorInformationUncertainty2001,hansenInverseProblemsNontrivial2012a,zuninoMonteCarloReservoir2015,hansenProbabilisticIntegrationGeoInformation2016a}. In the probabilistic approach, prior information is represented by a PDF $\rho(\mathbf{m})$ on the model parameters.

HMCLab provides a set of common PDFs for the prior, ranging from simple multivariate Gaussian distributions to more complex distributions such as a combination of Beta PDF-based marginals with a Gaussian copula to correlate the marginals. Another interesting prior is based on the Laplace distribution (related to the L1-norm) which promotes sparse (or blocky) models. Moreover, the user can provide his/her own prior by simply implementing functions with the appropriate signature (see the code documentation for more details). Any of the available priors can be combined with any of the available or user-generated forward models.

\section{Inferring complex information about the subsurface with HMCLab}

The HMCLab framework allows us to solve diverse inverse problems using sampling methods under a common platform. The software package includes a set of pre-defined geophysical forward and inverse problems, a set of prior distributions and allows the user to supply his/her own forward problem. In the following we show some examples of how to extract useful information about the subsurface for a set of selected geophysical inverse problems in the framework of the HMC method.

Once a collection of samples from the posterior distribution has been obtained, in order to calculate some arbitrary function $\phi(\mathbf{m})$ of $\mathbf{m}$, we can use the following relationship:
\begin{linenomath*}
  \begin{equation}
  \label{eq:mcmcpostanalysis}
  \int_{\rm M} \phi(\mathbf{m}) \sigma(\mathbf{m}) \, \mathrm{d} \mathbf{m} \approx \dfrac{1}{N} \sum_{i=1}^N \phi(\mathbf{m}_{i})
\end{equation}
\end{linenomath*}
where $N$ is the number of available samples and $\mathbf{m}_{i}$ represents one of the posterior models.

\subsection{2D full waveform acoustic inversion of reflection data}

\begin{figure}
  \includegraphics[width=\textwidth]{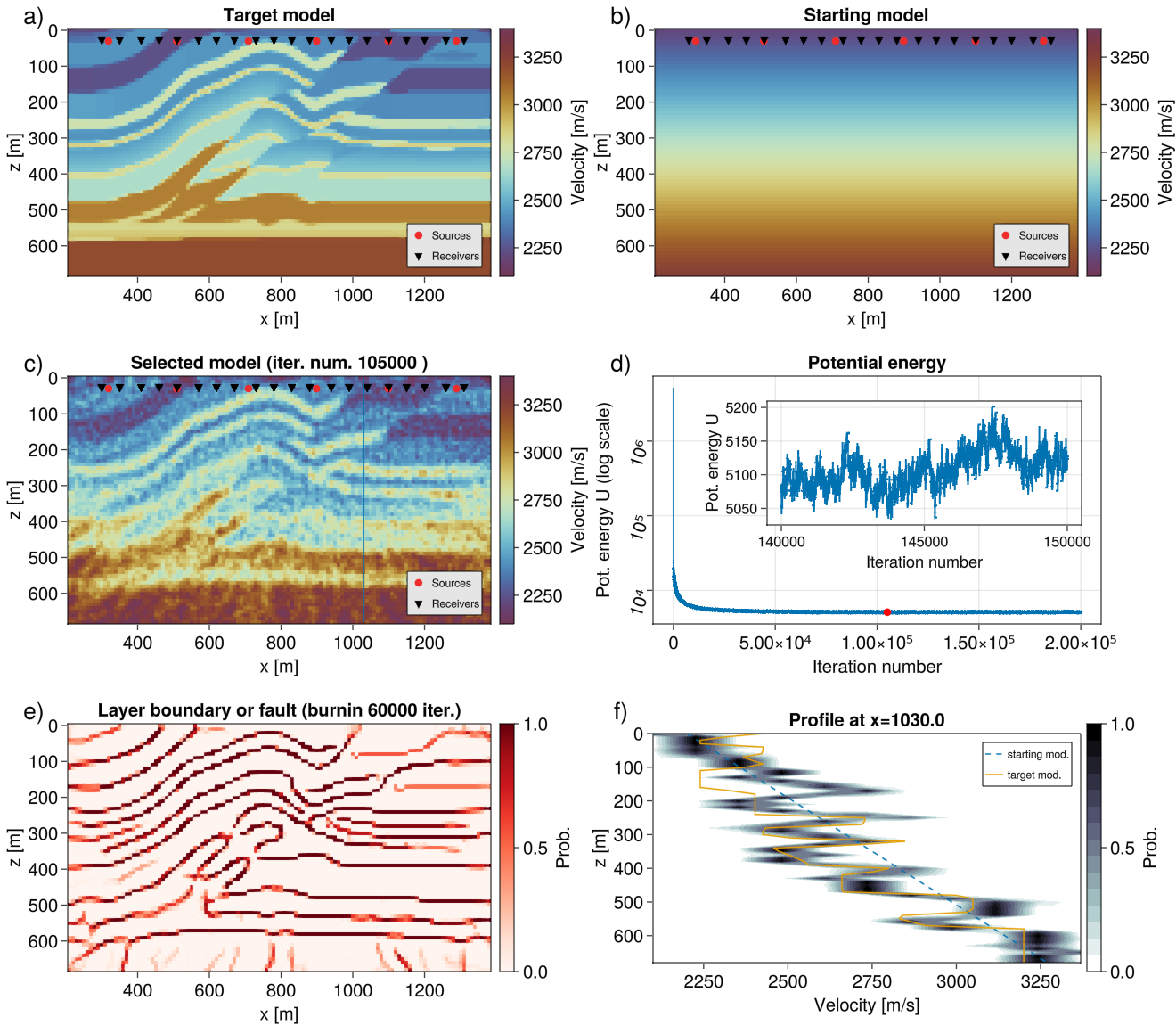}
  \caption{Acoustic waves inversion for velocity. a) Target  model of velocity, i.e., the one used to calculate the synthetic data. b) The starting velocity model used in the inversion. c) A randomly selected model from the collection of posterior models. d) A plot of the potential energy (misfit) as a function of iteration number, where the red dot indicates the potential energy of the model shown in panel d). e) A map of the probability of having a layer boundary of a fault computed using the collection of posterior models. f) A vertical profile of velocity showing the probability as computed from the collection of posterior models. The profile location is shown by a vertical line in panel c).}
  \label{fig:acousticprob}
\end{figure}

The first example is a 2D inversion of a seismic dataset based on the acoustic approximation. The forward problem is represented by the constant-density acoustic wave equation:
\begin{linenomath*}
  \begin{equation}
  \label{eq:acouwaveq}
  \dfrac{1}{v^2} \dfrac{\partial^2 u}{\partial t^2} = \dfrac{\partial^2 u}{\partial x^2} + \dfrac{\partial^2 u}{\partial z^2} + s
  \end{equation}
\end{linenomath*}
where $t$ is time, $x$ and $z$ the spatial coordinates, $u$ is the pressure field, $v$ the acoustic velocity and $s$ the source term. Forward calculations are carried out using a finite-difference scheme \citep{bunksMultiscaleSeismicWaveform1995,pasalicConvolutionalPerfectlyMatched2010}, where the model parameters are velocity at a set of grid points with size $(N_x \times N_z) = (160 \times 90)$ for the $x$- and $z$-direction respectively, for a total of $14400 $ model parameters. The grid spacing is 10 m in both directions. We assume the observational errors to be Gaussian distributed. Therefore, we use an $L_2$-norm potential energy function. The gradient of such a misfit function with respect to velocity is computed by means of the adjoint method for the acoustic wave equation, as described in \citet{bunksMultiscaleSeismicWaveform1995}. The use of the adjoint method enables us to efficiently evaluate the gradient, an essential prerequisite for being able to perform an HMC inversion. The geometry of the problem resembles the one typically found in exploration seismology, where active sources and receivers are located near the surface of the Earth, as shown in Fig.~\ref{fig:acousticprob}. The top boundary condition is a free surface, while the other sides are absorbing boundaries implemented as C-PML layers \citep{komatitschUnsplitConvolutionalPerfectly2007}. We use a set of 6 sources to generate synthetic data, add correlated Gaussian noise (standard deviation $0.05$, correlation length $0.01$ s) and use the result as the observed data to be inverted for the velocity model. To perform the inversion we use the NUTS algorithm \citep{hoffmanNoUturnSamplerAdaptively2014}, part of HMCLab. 

We ran $2 \times 10^5$ iterations of the NUTS algorithm, collecting about 45000 samples after thinning the chain and removing the models resulting from the burn-in phase. The starting velocity model is laterally homogeneous  (see Fig.~\ref{fig:acousticprob}b).The target model is a modified version of the SEG/EAGE overthrust model \citep{aminzadehSEGEAGE3D1997}. 
Fig.~\ref{fig:acousticprob}c shows a randomly chosen model from the collection of the posterior models. The model resembles the target model well, and all the different layers are visible.
The potential energy decreases rapidly within the first few hundreds of iterations, when the algorithm attempts to find a model which fits the large-scale structures (see Fig.~\ref{fig:acousticprob}d). Subsequently, the misfit keeps decreasing relatively slowly for much longer. We suspect this is due to the algorithm slowly adjusting the fine-scale structures, until it reaches a relatively stable misfit value. From the resulting collection of posterior models we can extract different pieces of information. One practical example is, e.g., calculating the probability of having a layer boundary or fault at any given node of the grid. To do so, we exploit eq.~\ref{eq:mcmcpostanalysis} using an indicator function $h(\mathbf{m})$ and compute  
\begin{linenomath*}
  \begin{equation}
  \label{eq:mcmcindicfun}
  \int_{\rm M} h(\mathbf{m}) \sigma(\mathbf{m}) \, \mathrm{d} \mathbf{m} \approx \dfrac{1}{N} \sum_{i=1}^N h(\mathbf{m}_{i})
\end{equation}
\end{linenomath*}
which produces a value of one in case a boundary/edge is detected and zero if not. The function $h(\mathbf{m})$ in this case is represented by a Canny edge detection filter \citep{cannyComputationalApproachEdge1986}, which is applied to each velocity model in the posterior collection, so that the model is transformed into a binary image of zeros and ones. Fig.~\ref{fig:acousticprob}e shows the results of such calculations. The large majority of the boundaries present in the target model appear as high probability structures in Fig.~\ref{fig:acousticprob}e, particularly at shallow depths. Finally, Fig.~\ref{fig:acousticprob}f shows a map of probability for a vertical profile of velocity at $x=1030.0$ m, showing the spread of the solutions. The profiles of the starting and target model are shown for comparison.

\subsection{First arrival traveltime hypocenter location using fiber-optic sensing}
An archetypal seismological inverse problem is the estimation of earthquake hypocenters in a medium with unknown structure and velocity using the first arrival times of the seicmic waves excited by the events. HMCLab supplies a simple approach to the hypocenter location problem that uses first arrival data, assuming a homogeneous medium. Arrival times are modelled by straight rays propagating through this homogeneous medium. The inverse problem only requires relative arrival times of a single phase.

Although the physics of this model seem simple, the strong trade-offs between location, origin time and medium velocity make it ideal for Bayesian inference methods. Especially in the presence of trade-offs, one would expect strongly correlated posteriors, for which the HMC algorithm works particularly well. 

\begin{figure}
  \includegraphics[width=\textwidth]{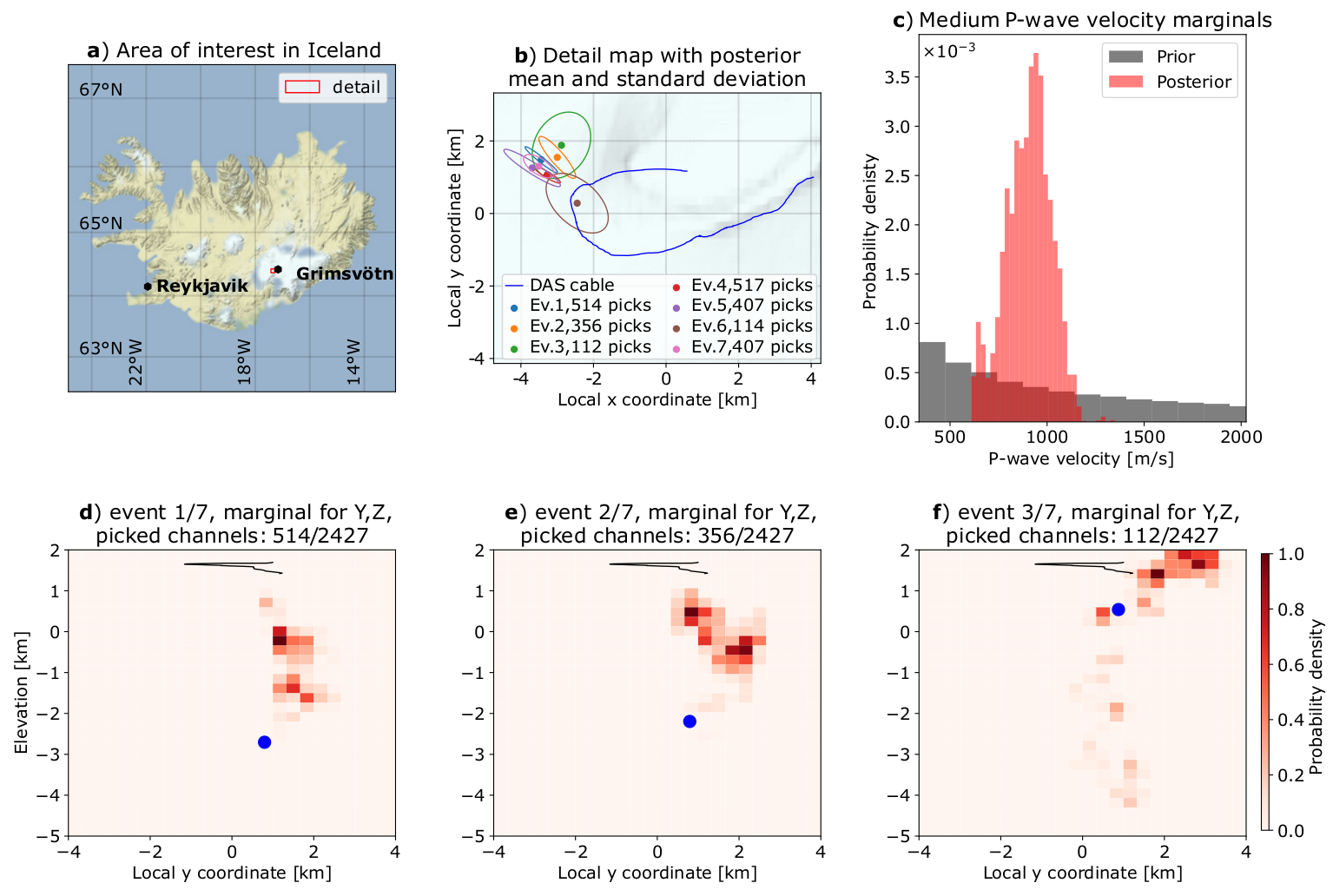}
  \caption{Source hypocenter location for data recorded on the Grimsv\"otn volcano. a) Overview of Iceland and location of the inset b). b) Geometry of the DAS acquisition fiber with posterior means and standard deviation ellipses oriented along principal axes. c) Marginal distribution of medium velocity prior and posterior to the inference. Note that the prior on medium velocity extends beyond the range of the plot. d)-f) Marginal distributions for the Y and Z components of the first 3 events. Note how for event 1 and 2, the volumes of uncertainty are more concentrated than for event 3.}
  \label{fig:triangulation}
\end{figure}

To illustrate how HMC performs on this inverse problem, we applied the algorithm to a dataset acquired on  Grimsvötn volcano in Iceland using a 12.5 km long Distributed Acoustic Sensing (DAS) fiber \citep{fichtnerFiberOpticObservation2022,klaasenSensingIcelandMost2022}, for which the acquisition geometry is shown in Fig.~\ref{fig:triangulation}.
Due to the use of DAS, this dataset has approximately 1500 separate channels.
We simultaneously infer the location of multiple events for which the effective medium velocity is assumed to be the same. The events are selected based on similarity in the observed move-out, as we expect these events to be relatively close.
By simulatenously inferring the location of multiple events the data better constrains the medium velocity, which reduces the trade-off between origin time and medium velocity compared to inferring location, origin time and medium velocity for a single event.
We define an $L_2$ misfit on the relative first arrivals of the picked phases, and only include those channels where the phase is picked. This means that some events have relatively less data points and therefore less importance within the inference. As prior, we use uniform distributions on location in a bounded cube of 20 km by 20 km by 10 km (width by length by depth) centered around the DAS cable. As the medium below the field site is unkown, we construct a prior on the P-wave velocity with a logarithmic uniform distribution (to take into account the fact that velocity is a positive parameter) between 340 and 7000 m/s, the extreme ends of possible medium material, i.e. air to relatively fast rock.

The results of a parallel tempered appraisal with 10 chains using HMC are given in Fig.~\ref{fig:triangulation}. The posterior on medium velocity, seen in Fig.~\ref{fig:triangulation}c, shows how, despite having a model with strong trade-offs in the parameters (namely origin time and medium P-wave velocity), one is still able to infer knowledge on one of these parameters, adding knowledge compared to the prior. Event 3, which was recorded on relatively few channels of the fiber-optic cable, features a high uncertainty of its location in the subsurface. This is in contrast to event 1 and 2, which seem to have a more concentrated volume of uncertainty. These results show that the posterior PDF of the location of these events is neither unimodal nor Gaussian, something which would have been difficult to deal with using deterministic methods. In general, one can see in Fig.~\ref{fig:triangulation}b that events with fewer picks are constrained less. Examples of relevant indicator functions~\citep{arnoldInterrogationTheory2018} for this inference would be the expected average depth of an event, or the probability of the medium velocity lying within a limited interval.

\subsection{First arrival tomography based on the eikonal equation}

\begin{figure}
  \includegraphics[width=\textwidth]{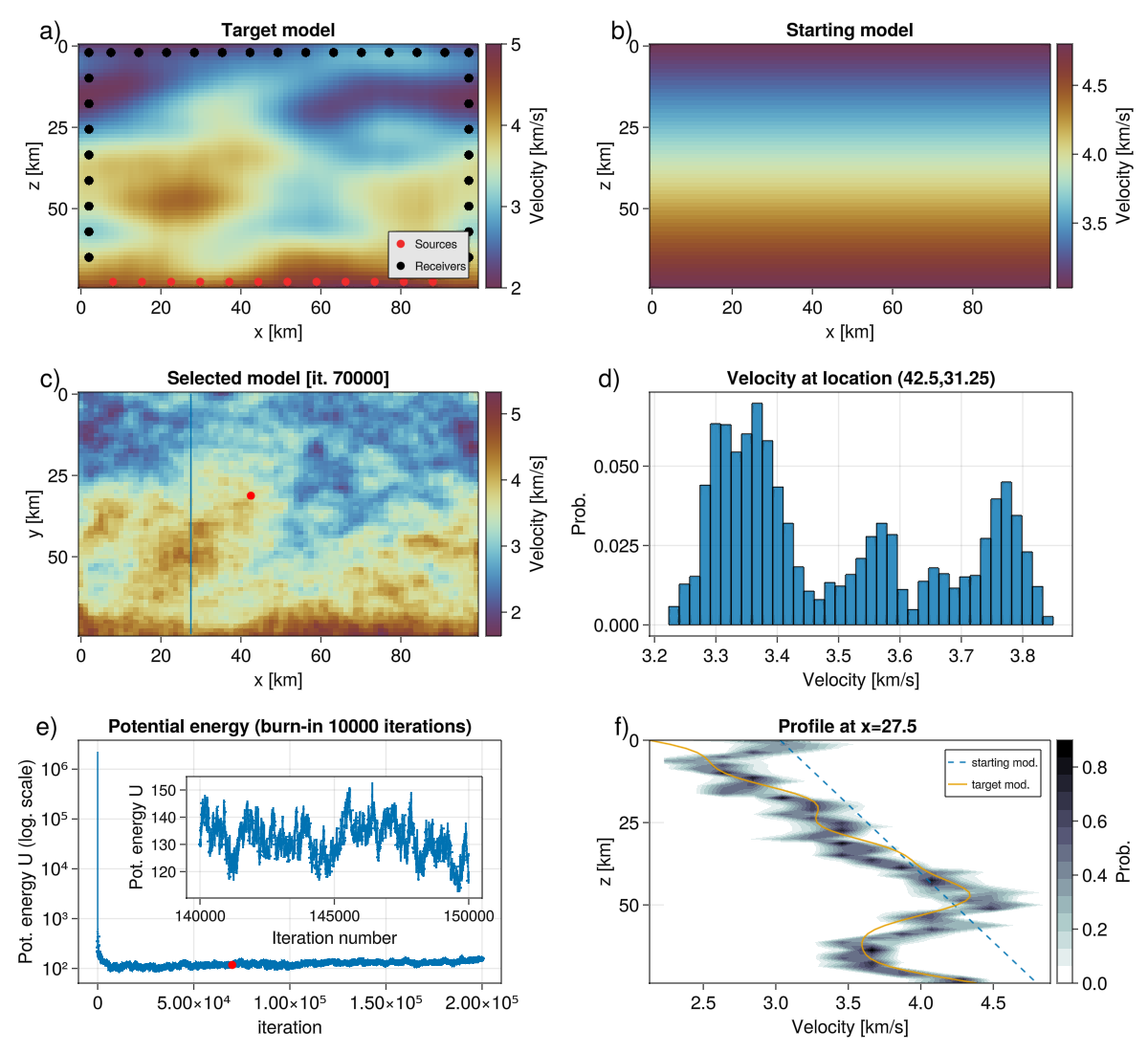}
  \caption{Nonlinear traveltime inversion. a) Target model used to generate the ``observed'' data, b) starting model, c) a selected model from the posterior collection, d) histogram of velocity at $(x,y) = (92,5,26.25)$ (marked by a red dot in c)), e) potential energy, f) profile of velocity showing a probability map as computed from the posterior collection of models (the profile location is shown by a vertical line in panel c)).}
  \label{fig:eikprob}
\end{figure}
Traveltime tomography is a popular approach for seismic inversion where the arrival time of seismic waves at given locations is used to infer the velocity structure of the subsurface. In this case, the forward model is represented by the eikonal equation:
\begin{linenomath*}
\begin{equation}
  \label{eq:eikforw}
 \sum_d \left( \dfrac{\partial \, t}{\partial x_d} \right)^2 = c^{-2}(x) \, ,
\end{equation}
\end{linenomath*}
where $t$ is the traveltime, $c$ the velocity and $d=2 \, \mathrm{ or } \, 3$ is the number of dimensions.
Since the forward model is nonlinear, typically ray paths are computed a priori in a reference Earth model and fixed, to linearize the problem and hence solve the inverse problem with gradient-based deterministic methods.
However, such strategy where a single solution is sought might miss some important information. Because of the nonlinearity of the problem, the misfit functional may feature multiple minima which cannot be detected with deterministic methods. On the contrary, a probabilistic approach may reveal different plausible velocity models to be consistent with the observed data, as we show in the following example.

The example we address here is to solve a 2D inverse problem with a geometry depicted in Fig.~\ref{fig:eikprob}, where we have a velocity model described by 4800 cells (80 in the $x$-direction and 60 in the $y$-direction), a set of sources randomly distributed close to the bottom of the model and a set of receivers near the surface and along the left and right sides of the model. The grid spacing is 1.25 km in both directions.
The forward problem is solved using a fast marching method (FMM) \cite[e.g.,][]{sethianFastMarchingLevel1996,rawlinsonWaveFrontEvolution2004,treisterFastMarchingAlgorithm2016} which computes the traveltimes at each point of a grid using a finite-difference strategy. The gradient of the Gaussian misfit functional with respect to velocity is computed by means of the adjoint method \cite[e.g.,][]{leungAdjointStateMethod2006,taillandierFirstarrivalTraveltimeTomography2009,zuninoIntegratingGradientInformation2018}.
To solve the inverse problem, we ran $2 \times 10^5$ iterations with the NUTS algorithm. The target model is shown in Fig.~\ref{fig:eikprob}a, while the starting model is depicted in Fig.~\ref{fig:eikprob}b. The latter is a laterally homogeneous velocity model. Panels c and d of Fig.~\ref{fig:eikprob} show a randomly selected model from the posterior collection and a histogram of the velocity at a given location as obtained by looking at all samples after the burn-in period. Interestingly, the histogram shows a multi-modal distribution, where probable velocity values cluster near three different values. This means that there are three different ranges of velocity values which are highly probable, i.e., they are all compatible with the observed data. Such finding would not be possible with an optimization method where the solution is represented by a single velocity model.
The potential energy (misfit) as a function of iterations (Fig.~\ref{fig:eikprob}) features a sharp decrease in the first hundreds of iterations and then a slow descent until equilibrium is reached around $5 \times 10^5$ iterations. Fig.~\ref{fig:eikprob}f shows a map of probability for a vertical profile at $x=27.5$ km, together with the profile for the starting and target model.

\subsection{Magnetic anomaly inversion with polygonal bodies}
\begin{figure}
  \includegraphics[width=\textwidth]{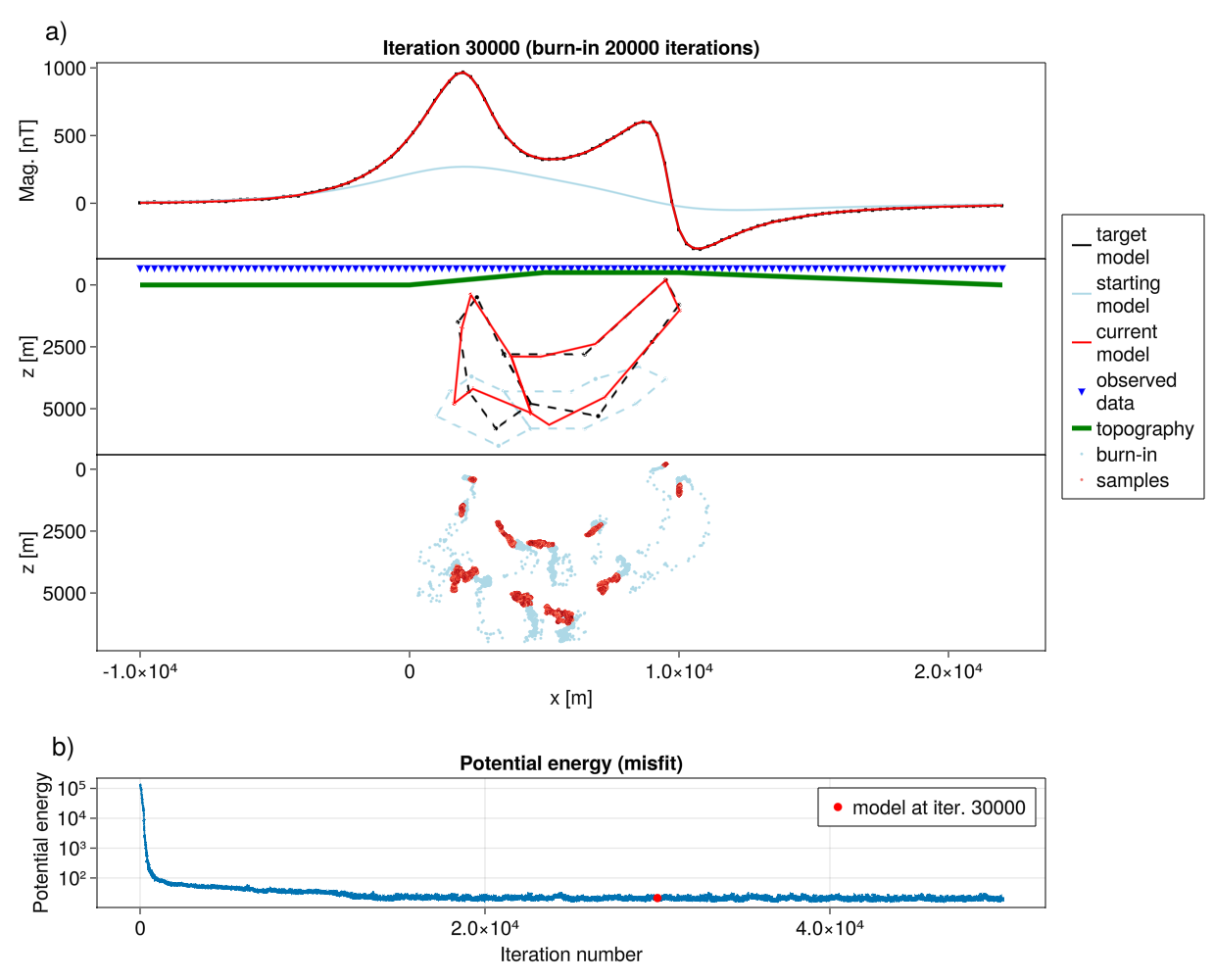}
  \caption{Magnetic anomaly problem. a) Three panels showing (from top to bottom): 1) the observed magnetic anomaly, the one calculated from the starting and selected models, 2) the starting, target and selected polygonal bodies including topography and position of measurements and 3) a scatter plot of the position of the vertices before and after the burn-in phase (every 10 iterations). b) A plot of the potential energy (log. scale) as a function of iteration number.}
  \label{fig:magprob}
\end{figure}
The last synthetic example presented is an inversion of magnetic anomaly data using a 2.75D parameterization in terms of polygonal bodies \cite[e.g.,][]{rasmussenEndCorrectionsPotential1979a,campbellBASICProgramsCalculate1983}. The design and detailed description of the method to construct and solve this inverse problem is the subject of another paper \citep{zuninoHamiltonianMonteCarlo2022}.
In this setup, each polygonal body has a homogeneous magnetization and its shape is controlled by the position of its vertices, which, in this example, are the unknowns of the inverse problem. In this setup, the relation between the position of vertices (model parameters) and the magnetic response is nonlinear. The label 2.75D means that the polygonal bodies have a given finite lateral extent in the $y$ direction that can be different in the $+y$ and $-y$ directions. 
The observed data are represented by a profile along the $x$ direction with about 130 observation points (see Fig.~\ref{fig:magprob}a). To solve the inverse problem, we ran $5 \times 10^4$ iterations using the NUTS algorithm. The potential energy decreases rapidly in the first few hundreds of iterations (see Fig.~\ref{fig:magprob}b) when the polygonal bodies move from the starting position to a more likely configuration, similar to the target model. Afterwards, the algorithm samples the posterior distribution, producing a set of posterior models. In this example the mass matrix contains non-zero off-diagonal elements in order to force the algorithm to avoid creating geologically implausible shapes. Such off-diagonal elements control the correlation of the momentum variables and hence indirectly the correlation shown in the models visited by the algorithm. 
A randomly selected model from the posterior collection is shown in  Fig.~\ref{fig:magprob}a, which fits the observed data well and is also close to the shape of the target model. In the last panel of Fig.~\ref{fig:magprob}a a set of position of the vertices before and after the burn-in phase are depicted, showing the relatively low uncertainty in the positions for the anomaly.

\section{HMCLab framework as a software package}

\begin{table} 
  \centering
  \begin{tabular}{l|c|c}
    \specialrule{0.1em}{0.25em}{0.25em}
    Inverse problem & Julia & Python \\ 
    \specialrule{0.1em}{0.25em}{0.25em}
    Traveltime tomography (eikonal eq., nonlinear) &  2D \& 3D (parall.) &  \\
    \specialrule{0.025em}{0.25em}{0.25em}
    Traveltime tomography (refracted rays, nonlinear) &   & 2D (parall.) \\\specialrule{0.025em}{0.25em}{0.25em}
    Traveltime tomography (straight rays, linear) &   & 2D (parall.) \\
    \specialrule{0.025em}{0.25em}{0.25em}
    Full-waveform inversion & 2D acoustic (parall.) & 2D elastic (parall.) \\
    \specialrule{0.025em}{0.25em}{0.25em}
    Earthquake source location & & 2D \& 3D \\
    \specialrule{0.025em}{0.25em}{0.25em}
    AVA seismic reflection data + rock physics & 1D, 2D, 3D (parall.) & \\
    \specialrule{0.025em}{0.25em}{0.25em}
    Magnetic anomalies (polygons) & 2D, 2.75D & \\
    \specialrule{0.025em}{0.25em}{0.25em}
    Gravity anomalies (polygons) & 2D,  2.75D & \\
    \specialrule{0.025em}{0.25em}{0.25em}
    Joint gravity and magnetic anomalies (polygons) & 2D, 2.75D & \\
    \specialrule{0.025em}{0.25em}{0.25em}
    Arbitrary full and sparse matrix equations& & \checkmark \\
    \specialrule{0.025em}{0.25em}{0.25em}
    User-provided forwards & \checkmark & \checkmark\\
                    & & \\
    \specialrule{0.1em}{0.25em}{0.25em}
    Prior distribution & Julia & Python \\ 
    \specialrule{0.1em}{0.25em}{0.25em}
    Gaussian (L2) & \checkmark & \checkmark \\
    \specialrule{0.025em}{0.25em}{0.25em}
    Laplace (L1) &  & \checkmark \\
    \specialrule{0.025em}{0.25em}{0.25em}
    Uniform & \checkmark & \checkmark \\
    \specialrule{0.025em}{0.25em}{0.25em}
    Beta marginals + Gaussian copula & \checkmark (parall.) &  \\
    \specialrule{0.025em}{0.25em}{0.25em}                                  
    Arbitrary mixture &  & \checkmark \\
    \specialrule{0.025em}{0.25em}{0.25em}
    User defined marginals & \checkmark & \checkmark \\
    \specialrule{0.025em}{0.25em}{0.25em}                                  
  \end{tabular}
  \vspace{0.2cm}
  \caption{Snapshot of currently available inverse problems and priors. HMCLab is constantly evolving, therefore changes are expected. 1-2-3D refers to the physical dimensions of the problem, "parall." means a parallelized (multi-core) implementation is available.}
  \label{tab:inverse_problems}
\end{table}
The HMCLab framework is practically implemented in a set of open-source software packages written in the Python \citep{vanrossumPythonTutorial1995} and Julia \citep{bezansonJuliaFreshApproach2017} programming languages. HMCLab is, in fact, a numerical laboratory to allow the user to experiment with sampling algorithms on different geophysical problems, ranging from purely educational examples to research-oriented studies. The aim is to provide a user-friendly framework where it is possible to experiment with various problems and algorithms and solve realistic inverse problems, either provided by HMCLab or created by the user.

Table~\ref{tab:inverse_problems} summarizes the geophysical problems and prior models which are currently available in HMCLab. Both Julia and Python implementations are modular in that, forward modelling, gradient calculations and sampler (or optimizer) can be combined arbitrarily. Moreover, in addition to the listed problems, the sampler (or optimizer) can be used on user-defined problems. 
The main categories of inverse problems currently addressed by HMCLab are:
\begin{itemize}
\item linearized (straight rays) and nonlinear (eikonal equation) traveltime tomography,
\item full waveform inversion in 2D in the acoustic and elastic formulations (P-SV),
\item earthquake source location in 3D based on the straight-ray approximation,
\item joint (or independent) gravity and magnetic anomaly inversion in 2.75D using polygonal bodies,
\item amplitude versus angle (AVA) seismic data including a rock physics model in 3D.
\end{itemize}
In addition, a set of priors is provided, which can be combined with any problem. For all the above mentioned geophysical problems, HMCLab provides functions to solve forward problems and to compute the gradient of misfit functions with respect to model parameters. For all these physics, samplers are available to appraise the inverse problem. However, the functions of these physics can be used independently of the sampler, hence the user can construct his/her own inversion scheme. Also, the user may supply his/her own forward model, prior and gradient calculation code and subsequently use the available inversion algorithms by providing a minimum set of functions with the appropriate signature as described in the documentation.

HMCLab is not limited to flavors of the HMC algorithm, but includes other more traditional algorithms such as the random walk Metropolis-Hastings algorithm \citep{metropolisEquationStateCalculations1953a, hastingsMonteCarloSampling1970}, where gradients are not used. Moreover, we provide an interface \citep{gebraadSimpleSVGDTinyInterface2022} to the Stein Variational Gradient Descent (SVGD, \cite{liuSteinVariationalGradient2019}) variational inference algorithm, providing an alternative probabilistic appraisal algorithm to the included MCMC algorithms.
As already mentioned, HMCLab includes functions to compute gradients of the misfit functional, and, as such, deterministic inversions are also possible \cite[e.g.,][]{zuninoEfficientMethodSolve2019}. An example is basic gradient descent, where the modes (local minima) of the defined posterior distribution can be found deterministically. As such, HMCLab also facilitates the usage of Python \cite[e.g.,][] {virtanenSciPyFundamentalAlgorithms2020} and Julia optimization libraries, including popular algorithms such as the (Limited Memory) Broyden-Fletcher-Goldfarb-Shanno and Newton Conjugate Gradient algorithms \citep{nocedalNumericalOptimization2006}.

Noteworthy is the inclusion of several notebooks that illustrate the basic concepts of MCMC sampling in general, applied to HMCLab in particular. This ranges from investigating the basic properties of a Markov chain, such as number of proposals, stepsizes and resulting acceptance rates, to tuning the various included algorithms and even implementing one's own inverse problems. These notebooks are available to all users to run out-of-the-box in our supplied Docker environment, but are also available without a Python or Julia interpreter as plain HTML. HMCLab homepage can be found at \url{https://hmclab.science}, while the Julia and Python versions at \url{https://gitlab.com/JuliaGeoph} and \url{https://python.hmclab.science/index.html}, respectively.
Finally, HMCLab is constantly updated and expanded and contributions from the community are welcomed.

\section{Conclusions}
In this work we have shown how a common framework for probabilistic inversion can be utilized to solve a diverse range of geophysical problems. In particular, the HMC method can be used to solve a variety of nonlinear geophysical inverse problems. In contrast to other MCMC algorithms, HMC exploits the information derived from the gradient of the posterior pdf to drive the sampling towards regions of high probability, hence being able to traverse the model space more efficiently. This property is very beneficial, especially for problems which allow efficient computation of the gradients, e.g., when using the adjoint method, analytical derivatives or automatic differentiation.
In this paper we have shown a set of example problems including full-waveform acoustic inversion, hypocenter location, traveltime tomography and magnetic anomaly inversion using polygonal bodies. The examples presented have been solved using the software implementation HMCLab, a numerical laboratory for better understanding and solving inverse problems in a probabilistic manner, written in the high-level languages Julia and Python. By providing a collection of models as the solution of the inverse problem, HMCLab enables the user to perform a statistical analysis and retrieve desired probabilistic information. HMCLab is in constant evolution, and hopefully will be augmented by contributions from interested users. In addition to the available forward problems and priors, we made it accessible to the user to construct their own inverse problem and easily apply the methods provided by HMCLab. Moreover, several types of prior information are available, allowing the user to adequately describe prior certainties and uncertainties.

\section*{Acknowledgments}
($\dagger$) The first two authors contributed equally to this work.

We wish to thank Klaus Mosegaard, who, over the years, has contributed much to the authors' understanding of Markov chain Monte Carlo methods and their use in Geosciences. We also would like to thank Sara Klaasen for sharing the Grimsvötn DAS dataset. L.\ G.\ would like to thank Xin Zhang and Andrew Curtis for fruitful discussions on Bayesian inference and Stein Variational Gradient Descent. Furthermore, thanks go out to anyone who dared to test the software in the experimental state before this publication to help us improve and debug HMCLab; Cyrill, Marc, Runa, Sara, Ariane, and to anyone who will contribute in the future.
This work has been supported by the Swiss National Science Foundation (``Hamiltonian Monte Carlo Full-Waveform
Inversion'', grant number 200021\_192236).

\bibliographystyle{apacite}
\bibliography{HMCLab_bibliography}

\begin{thebibliography}{}

\bibitem [\protect \citeauthoryear {%
Aminzadeh%
\ \BBA {} Brac%
}{%
Aminzadeh%
\ \BBA {} Brac%
}{%
{\protect \APACyear {1997}}%
}]{%
aminzadehSEGEAGE3D1997}
\APACinsertmetastar {%
aminzadehSEGEAGE3D1997}%
\begin{APACrefauthors}%
Aminzadeh, F.%
\BCBT {}\ \BBA {} Brac, J.%
\end{APACrefauthors}%
\unskip\
\newblock
\APACrefYearMonthDay{1997}{{\APACmonth{01}}}{}.
\newblock
\APACrefbtitle {{{SEG}}/{{EAGE}} 3-{{D Overthrust Models}}.} {{{SEG}}/{{EAGE}}
  3-{{D Overthrust Models}}.}
\newblock
\APACaddressPublisher{}{{Zenodo}}.
\newblock
\begin{APACrefDOI} \doi{10.5281/zenodo.4252588} \end{APACrefDOI}
\PrintBackRefs{\CurrentBib}

\bibitem [\protect \citeauthoryear {%
Arnold%
\ \BBA {} Curtis%
}{%
Arnold%
\ \BBA {} Curtis%
}{%
{\protect \APACyear {2018}}%
}]{%
arnoldInterrogationTheory2018}
\APACinsertmetastar {%
arnoldInterrogationTheory2018}%
\begin{APACrefauthors}%
Arnold, R.%
\BCBT {}\ \BBA {} Curtis, A.%
\end{APACrefauthors}%
\unskip\
\newblock
\APACrefYearMonthDay{2018}{{\APACmonth{09}}}{}.
\newblock
{\BBOQ}\APACrefatitle {Interrogation Theory} {Interrogation theory}.{\BBCQ}
\newblock
\APACjournalVolNumPages{Geophysical Journal International}{214}{3}{1830--1846}.
\newblock
\begin{APACrefDOI} \doi{10.1093/gji/ggy248} \end{APACrefDOI}
\PrintBackRefs{\CurrentBib}

\bibitem [\protect \citeauthoryear {%
Backus%
\ \BBA {} Gilbert%
}{%
Backus%
\ \BBA {} Gilbert%
}{%
{\protect \APACyear {1967}}%
}]{%
backusNumericalApplicationsFormalism1967a}
\APACinsertmetastar {%
backusNumericalApplicationsFormalism1967a}%
\begin{APACrefauthors}%
Backus, G\BPBI E.%
\BCBT {}\ \BBA {} Gilbert, J\BPBI F.%
\end{APACrefauthors}%
\unskip\
\newblock
\APACrefYearMonthDay{1967}{{\APACmonth{07}}}{}.
\newblock
{\BBOQ}\APACrefatitle {Numerical {{Applications}} of a {{Formalism}} for
  {{Geophysical Inverse Problems}}} {Numerical {{Applications}} of a
  {{Formalism}} for {{Geophysical Inverse Problems}}}.{\BBCQ}
\newblock
\APACjournalVolNumPages{Geophysical Journal International}{13}{1-3}{247--276}.
\newblock
\begin{APACrefDOI} \doi{10.1111/j.1365-246X.1967.tb02159.x} \end{APACrefDOI}
\PrintBackRefs{\CurrentBib}

\bibitem [\protect \citeauthoryear {%
Bezanson%
, Edelman%
, Karpinski%
\BCBL {}\ \BBA {} Shah%
}{%
Bezanson%
\ \protect \BOthers {.}}{%
{\protect \APACyear {2017}}%
}]{%
bezansonJuliaFreshApproach2017}
\APACinsertmetastar {%
bezansonJuliaFreshApproach2017}%
\begin{APACrefauthors}%
Bezanson, J.%
, Edelman, A.%
, Karpinski, S.%
\BCBL {}\ \BBA {} Shah, V\BPBI B.%
\end{APACrefauthors}%
\unskip\
\newblock
\APACrefYearMonthDay{2017}{}{}.
\newblock
{\BBOQ}\APACrefatitle {Julia: {{A Fresh Approach}} to {{Numerical Computing}}}
  {Julia: {{A Fresh Approach}} to {{Numerical Computing}}}.{\BBCQ}
\newblock
\APACjournalVolNumPages{SIAM Review}{59}{1}{65--98}.
\newblock
\begin{APACrefDOI} \doi{10.1137/141000671} \end{APACrefDOI}
\PrintBackRefs{\CurrentBib}

\bibitem [\protect \citeauthoryear {%
Bodin%
\ \BBA {} Sambridge%
}{%
Bodin%
\ \BBA {} Sambridge%
}{%
{\protect \APACyear {2009}}%
}]{%
bodinSeismicTomographyReversible2009}
\APACinsertmetastar {%
bodinSeismicTomographyReversible2009}%
\begin{APACrefauthors}%
Bodin, T.%
\BCBT {}\ \BBA {} Sambridge, M.%
\end{APACrefauthors}%
\unskip\
\newblock
\APACrefYearMonthDay{2009}{{\APACmonth{09}}}{}.
\newblock
{\BBOQ}\APACrefatitle {Seismic Tomography with the Reversible Jump Algorithm}
  {Seismic tomography with the reversible jump algorithm}.{\BBCQ}
\newblock
\APACjournalVolNumPages{Geophysical Journal International}{178}{3}{1411--1436}.
\newblock
\begin{APACrefDOI} \doi{10.1111/j.1365-246X.2009.04226.x} \end{APACrefDOI}
\PrintBackRefs{\CurrentBib}

\bibitem [\protect \citeauthoryear {%
Bosch%
}{%
Bosch%
}{%
{\protect \APACyear {1999}}%
}]{%
boschLithologicTomographyPlural1999}
\APACinsertmetastar {%
boschLithologicTomographyPlural1999}%
\begin{APACrefauthors}%
Bosch, M.%
\end{APACrefauthors}%
\unskip\
\newblock
\APACrefYearMonthDay{1999}{}{}.
\newblock
{\BBOQ}\APACrefatitle {Lithologic Tomography: {{From}} Plural Geophysical Data
  to Lithology Estimation} {Lithologic tomography: {{From}} plural geophysical
  data to lithology estimation}.{\BBCQ}
\newblock
\APACjournalVolNumPages{Journal of Geophysical Research: Solid
  Earth}{104}{B1}{749--766}.
\newblock
\begin{APACrefDOI} \doi{10.1029/1998JB900014} \end{APACrefDOI}
\PrintBackRefs{\CurrentBib}

\bibitem [\protect \citeauthoryear {%
Bunks%
, Saleck%
, Zaleski%
\BCBL {}\ \BBA {} Chavent%
}{%
Bunks%
\ \protect \BOthers {.}}{%
{\protect \APACyear {1995}}%
}]{%
bunksMultiscaleSeismicWaveform1995}
\APACinsertmetastar {%
bunksMultiscaleSeismicWaveform1995}%
\begin{APACrefauthors}%
Bunks, C.%
, Saleck, F\BPBI M.%
, Zaleski, S.%
\BCBL {}\ \BBA {} Chavent, G.%
\end{APACrefauthors}%
\unskip\
\newblock
\APACrefYearMonthDay{1995}{{\APACmonth{10}}}{}.
\newblock
{\BBOQ}\APACrefatitle {Multiscale Seismic Waveform Inversion} {Multiscale
  seismic waveform inversion}.{\BBCQ}
\newblock
\APACjournalVolNumPages{Geophysics}{60}{5}{1457--1473}.
\newblock
\begin{APACrefDOI} \doi{10.1190/1.1443880} \end{APACrefDOI}
\PrintBackRefs{\CurrentBib}

\bibitem [\protect \citeauthoryear {%
Campbell%
}{%
Campbell%
}{%
{\protect \APACyear {1983}}%
}]{%
campbellBASICProgramsCalculate1983}
\APACinsertmetastar {%
campbellBASICProgramsCalculate1983}%
\begin{APACrefauthors}%
Campbell, D\BPBI L.%
\end{APACrefauthors}%
\unskip\
\newblock
\APACrefYearMonthDay{1983}{}{}.
\newblock
\APACrefbtitle {{{BASIC}} Programs to Calculate Gravity and Magnetic Anomalies
  for 2 1/2 - Dimensional Prismatic Bodies} {{{BASIC}} programs to calculate
  gravity and magnetic anomalies for 2 1/2 - dimensional prismatic bodies}\
  \APACbVolEdTR{}{\BTR{}\ \BNUM\ 83-154}.
\newblock
\APACaddressInstitution{}{{U.S. Geological Survey,}}.
\newblock
\begin{APACrefDOI} \doi{10.3133/ofr83154} \end{APACrefDOI}
\PrintBackRefs{\CurrentBib}

\bibitem [\protect \citeauthoryear {%
Canny%
}{%
Canny%
}{%
{\protect \APACyear {1986}}%
}]{%
cannyComputationalApproachEdge1986}
\APACinsertmetastar {%
cannyComputationalApproachEdge1986}%
\begin{APACrefauthors}%
Canny, J.%
\end{APACrefauthors}%
\unskip\
\newblock
\APACrefYearMonthDay{1986}{{\APACmonth{11}}}{}.
\newblock
{\BBOQ}\APACrefatitle {A {{Computational Approach}} to {{Edge Detection}}} {A
  {{Computational Approach}} to {{Edge Detection}}}.{\BBCQ}
\newblock
\APACjournalVolNumPages{IEEE Transactions on Pattern Analysis and Machine
  Intelligence}{PAMI-8}{6}{679--698}.
\newblock
\begin{APACrefDOI} \doi{10.1109/TPAMI.1986.4767851} \end{APACrefDOI}
\PrintBackRefs{\CurrentBib}

\bibitem [\protect \citeauthoryear {%
Creutz%
}{%
Creutz%
}{%
{\protect \APACyear {1988}}%
}]{%
creutzGlobalMonteCarlo1988}
\APACinsertmetastar {%
creutzGlobalMonteCarlo1988}%
\begin{APACrefauthors}%
Creutz, M.%
\end{APACrefauthors}%
\unskip\
\newblock
\APACrefYearMonthDay{1988}{{\APACmonth{08}}}{}.
\newblock
{\BBOQ}\APACrefatitle {Global {{Monte Carlo}} Algorithms for Many-Fermion
  Systems} {Global {{Monte Carlo}} algorithms for many-fermion systems}.{\BBCQ}
\newblock
\APACjournalVolNumPages{Physical Review D}{38}{4}{1228--1238}.
\newblock
\begin{APACrefDOI} \doi{10.1103/PhysRevD.38.1228} \end{APACrefDOI}
\PrintBackRefs{\CurrentBib}

\bibitem [\protect \citeauthoryear {%
Curtis%
\ \BBA {} Lomax%
}{%
Curtis%
\ \BBA {} Lomax%
}{%
{\protect \APACyear {2001}}%
{\protect \APACexlab {{\protect \BCnt {1}}}}}]{%
curtisPriorInformationSampling2001}
\APACinsertmetastar {%
curtisPriorInformationSampling2001}%
\begin{APACrefauthors}%
Curtis, A.%
\BCBT {}\ \BBA {} Lomax, A.%
\end{APACrefauthors}%
\unskip\
\newblock
\APACrefYearMonthDay{2001{\protect \BCnt {1}}}{}{}.
\newblock
{\BBOQ}\APACrefatitle {Prior Information, Sampling Distributions, and the Curse
  of Dimensionality} {Prior information, sampling distributions, and the curse
  of dimensionality}.{\BBCQ}
\newblock
\APACjournalVolNumPages{Geophysics}{66}{2}{372--378}.
\PrintBackRefs{\CurrentBib}

\bibitem [\protect \citeauthoryear {%
Curtis%
\ \BBA {} Lomax%
}{%
Curtis%
\ \BBA {} Lomax%
}{%
{\protect \APACyear {2001}}%
{\protect \APACexlab {{\protect \BCnt {2}}}}}]{%
curtisPriorInformationSampling2001a}
\APACinsertmetastar {%
curtisPriorInformationSampling2001a}%
\begin{APACrefauthors}%
Curtis, A.%
\BCBT {}\ \BBA {} Lomax, A.%
\end{APACrefauthors}%
\unskip\
\newblock
\APACrefYearMonthDay{2001{\protect \BCnt {2}}}{{\APACmonth{03}}}{}.
\newblock
{\BBOQ}\APACrefatitle {Prior Information, Sampling Distributions, and the Curse
  of Dimensionality} {Prior information, sampling distributions, and the curse
  of dimensionality}.{\BBCQ}
\newblock
\APACjournalVolNumPages{Geophysics}{66}{}{372--378}.
\newblock
\begin{APACrefDOI} \doi{10.1190/1.1444928} \end{APACrefDOI}
\PrintBackRefs{\CurrentBib}

\bibitem [\protect \citeauthoryear {%
D{\k{e}}bski%
}{%
D{\k{e}}bski%
}{%
{\protect \APACyear {2010}}%
}]{%
debskiChapterProbabilisticInverse2010}
\APACinsertmetastar {%
debskiChapterProbabilisticInverse2010}%
\begin{APACrefauthors}%
D{\k{e}}bski, W.%
\end{APACrefauthors}%
\unskip\
\newblock
\APACrefYearMonthDay{2010}{{\APACmonth{01}}}{}.
\newblock
{\BBOQ}\APACrefatitle {Chapter 1 - {{Probabilistic Inverse Theory}}} {Chapter 1
  - {{Probabilistic Inverse Theory}}}.{\BBCQ}
\newblock
\BIn{} R.~Dmowska\ (\BED), \APACrefbtitle {Advances in {{Geophysics}}}
  {Advances in {{Geophysics}}}\ (\BVOL~52, \BPGS\ 1--102).
\newblock
\APACaddressPublisher{}{{Elsevier}}.
\newblock
\begin{APACrefDOI} \doi{10.1016/S0065-2687(10)52001-6} \end{APACrefDOI}
\PrintBackRefs{\CurrentBib}

\bibitem [\protect \citeauthoryear {%
Duane%
, Kennedy%
, Pendleton%
\BCBL {}\ \BBA {} Roweth%
}{%
Duane%
\ \protect \BOthers {.}}{%
{\protect \APACyear {1987}}%
}]{%
duaneHybridMonteCarlo1987}
\APACinsertmetastar {%
duaneHybridMonteCarlo1987}%
\begin{APACrefauthors}%
Duane, S.%
, Kennedy, A\BPBI D.%
, Pendleton, B\BPBI J.%
\BCBL {}\ \BBA {} Roweth, D.%
\end{APACrefauthors}%
\unskip\
\newblock
\APACrefYearMonthDay{1987}{}{}.
\newblock
{\BBOQ}\APACrefatitle {Hybrid {{Monte Carlo}}} {Hybrid {{Monte Carlo}}}.{\BBCQ}
\newblock
\APACjournalVolNumPages{Physics Letters B}{195}{2}{216--222}.
\newblock
\begin{APACrefDOI} \doi{10.1016/0370-2693(87)91197-X} \end{APACrefDOI}
\PrintBackRefs{\CurrentBib}

\bibitem [\protect \citeauthoryear {%
Fichtner%
, Bunge%
\BCBL {}\ \BBA {} Igel%
}{%
Fichtner%
\ \protect \BOthers {.}}{%
{\protect \APACyear {2006}}%
}]{%
fichtnerAdjointMethodSeismology2006a}
\APACinsertmetastar {%
fichtnerAdjointMethodSeismology2006a}%
\begin{APACrefauthors}%
Fichtner, A.%
, Bunge, H\BPBI P.%
\BCBL {}\ \BBA {} Igel, H.%
\end{APACrefauthors}%
\unskip\
\newblock
\APACrefYearMonthDay{2006}{{\APACmonth{08}}}{}.
\newblock
{\BBOQ}\APACrefatitle {The Adjoint Method in Seismology: {{I}}. {{Theory}}}
  {The adjoint method in seismology: {{I}}. {{Theory}}}.{\BBCQ}
\newblock
\APACjournalVolNumPages{Physics of the Earth and Planetary
  Interiors}{157}{1}{86--104}.
\newblock
\begin{APACrefDOI} \doi{10.1016/j.pepi.2006.03.016} \end{APACrefDOI}
\PrintBackRefs{\CurrentBib}

\bibitem [\protect \citeauthoryear {%
Fichtner%
\ \protect \BOthers {.}}{%
Fichtner%
\ \protect \BOthers {.}}{%
{\protect \APACyear {2022}}%
}]{%
fichtnerFiberOpticObservation2022}
\APACinsertmetastar {%
fichtnerFiberOpticObservation2022}%
\begin{APACrefauthors}%
Fichtner, A.%
, Klaasen, S.%
, Thrastarson, S.%
, {\c C}ubuk-Sabuncu, Y.%
, Paitz, P.%
\BCBL {}\ \BBA {} J{\'o}nsd{\'o}ttir, K.%
\end{APACrefauthors}%
\unskip\
\newblock
\APACrefYearMonthDay{2022}{{\APACmonth{07}}}{}.
\newblock
{\BBOQ}\APACrefatitle {Fiber-{{Optic Observation}} of {{Volcanic Tremor}}
  through {{Floating Ice Sheet Resonance}}} {Fiber-{{Optic Observation}} of
  {{Volcanic Tremor}} through {{Floating Ice Sheet Resonance}}}.{\BBCQ}
\newblock
\APACjournalVolNumPages{The Seismic Record}{2}{3}{148--155}.
\newblock
\begin{APACrefDOI} \doi{10.1785/0320220010} \end{APACrefDOI}
\PrintBackRefs{\CurrentBib}

\bibitem [\protect \citeauthoryear {%
Fichtner%
, Zunino%
\BCBL {}\ \BBA {} Gebraad%
}{%
Fichtner%
\ \protect \BOthers {.}}{%
{\protect \APACyear {2019}}%
}]{%
fichtnerHamiltonianMonteCarlo2019}
\APACinsertmetastar {%
fichtnerHamiltonianMonteCarlo2019}%
\begin{APACrefauthors}%
Fichtner, A.%
, Zunino, A.%
\BCBL {}\ \BBA {} Gebraad, L.%
\end{APACrefauthors}%
\unskip\
\newblock
\APACrefYearMonthDay{2019}{}{}.
\newblock
{\BBOQ}\APACrefatitle {Hamiltonian {{Monte Carlo}} Solution of Tomographic
  Inverse Problems} {Hamiltonian {{Monte Carlo}} solution of tomographic
  inverse problems}.{\BBCQ}
\newblock
\APACjournalVolNumPages{Geophysical Journal International}{216}{2}{1344--1363}.
\newblock
\begin{APACrefDOI} \doi{10.1093/gji/ggy496} \end{APACrefDOI}
\PrintBackRefs{\CurrentBib}

\bibitem [\protect \citeauthoryear {%
Fichtner%
, Zunino%
, Gebraad%
\BCBL {}\ \BBA {} Boehm%
}{%
Fichtner%
\ \protect \BOthers {.}}{%
{\protect \APACyear {2021}}%
}]{%
fichtnerAutotuningHamiltonianMonte2021}
\APACinsertmetastar {%
fichtnerAutotuningHamiltonianMonte2021}%
\begin{APACrefauthors}%
Fichtner, A.%
, Zunino, A.%
, Gebraad, L.%
\BCBL {}\ \BBA {} Boehm, C.%
\end{APACrefauthors}%
\unskip\
\newblock
\APACrefYearMonthDay{2021}{}{}.
\newblock
{\BBOQ}\APACrefatitle {Autotuning {{Hamiltonian Monte Carlo}} for Efficient
  Generalized Nullspace Exploration} {Autotuning {{Hamiltonian Monte Carlo}}
  for efficient generalized nullspace exploration}.{\BBCQ}
\newblock
\APACjournalVolNumPages{Geophysical Journal International}{227}{2}{941--968}.
\newblock
\begin{APACrefDOI} \doi{10.1093/gji/ggab270} \end{APACrefDOI}
\PrintBackRefs{\CurrentBib}

\bibitem [\protect \citeauthoryear {%
Gebraad%
}{%
Gebraad%
}{%
{\protect \APACyear {2022}}%
}]{%
gebraadSimpleSVGDTinyInterface2022}
\APACinsertmetastar {%
gebraadSimpleSVGDTinyInterface2022}%
\begin{APACrefauthors}%
Gebraad, L.%
\end{APACrefauthors}%
\unskip\
\newblock
\APACrefYearMonthDay{2022}{{\APACmonth{01}}}{}.
\newblock
\APACrefbtitle {{{simpleSVGD}}: {{A}} Tiny Interface to {{Stein Variational
  Gradient Descent}} Using Various Optimization Algorithms.} {{{simpleSVGD}}:
  {{A}} tiny interface to {{Stein Variational Gradient Descent}} using various
  optimization algorithms.}
\newblock
\APAChowpublished {Zenodo}.
\newblock
\begin{APACrefDOI} \doi{10.5281/zenodo.5938430} \end{APACrefDOI}
\PrintBackRefs{\CurrentBib}

\bibitem [\protect \citeauthoryear {%
Gebraad%
, Boehm%
\BCBL {}\ \BBA {} Fichtner%
}{%
Gebraad%
\ \protect \BOthers {.}}{%
{\protect \APACyear {2020}}%
}]{%
gebraadBayesianElasticFullWaveform2020}
\APACinsertmetastar {%
gebraadBayesianElasticFullWaveform2020}%
\begin{APACrefauthors}%
Gebraad, L.%
, Boehm, C.%
\BCBL {}\ \BBA {} Fichtner, A.%
\end{APACrefauthors}%
\unskip\
\newblock
\APACrefYearMonthDay{2020}{}{}.
\newblock
{\BBOQ}\APACrefatitle {Bayesian {{Elastic Full-Waveform Inversion Using
  Hamiltonian Monte Carlo}}} {Bayesian {{Elastic Full-Waveform Inversion Using
  Hamiltonian Monte Carlo}}}.{\BBCQ}
\newblock
\APACjournalVolNumPages{Journal of Geophysical Research: Solid
  Earth}{125}{3}{e2019JB018428}.
\newblock
\begin{APACrefDOI} \doi{10.1029/2019JB018428} \end{APACrefDOI}
\PrintBackRefs{\CurrentBib}

\bibitem [\protect \citeauthoryear {%
Geman%
\ \BBA {} Geman%
}{%
Geman%
\ \BBA {} Geman%
}{%
{\protect \APACyear {1984}}%
}]{%
gemanStochasticRelaxationGibbs1984}
\APACinsertmetastar {%
gemanStochasticRelaxationGibbs1984}%
\begin{APACrefauthors}%
Geman, S.%
\BCBT {}\ \BBA {} Geman, D.%
\end{APACrefauthors}%
\unskip\
\newblock
\APACrefYearMonthDay{1984}{{\APACmonth{11}}}{}.
\newblock
{\BBOQ}\APACrefatitle {Stochastic {{Relaxation}}, {{Gibbs Distributions}}, and
  the {{Bayesian Restoration}} of {{Images}}} {Stochastic {{Relaxation}},
  {{Gibbs Distributions}}, and the {{Bayesian Restoration}} of
  {{Images}}}.{\BBCQ}
\newblock
\APACjournalVolNumPages{IEEE Transactions on Pattern Analysis and Machine
  Intelligence}{PAMI-6}{6}{721--741}.
\newblock
\begin{APACrefDOI} \doi{10.1109/TPAMI.1984.4767596} \end{APACrefDOI}
\PrintBackRefs{\CurrentBib}

\bibitem [\protect \citeauthoryear {%
Gilks%
, Best%
\BCBL {}\ \BBA {} Tan%
}{%
Gilks%
\ \protect \BOthers {.}}{%
{\protect \APACyear {1995}}%
}]{%
gilksAdaptiveRejectionMetropolis1995}
\APACinsertmetastar {%
gilksAdaptiveRejectionMetropolis1995}%
\begin{APACrefauthors}%
Gilks, W\BPBI R.%
, Best, N\BPBI G.%
\BCBL {}\ \BBA {} Tan, K\BPBI K\BPBI C.%
\end{APACrefauthors}%
\unskip\
\newblock
\APACrefYearMonthDay{1995}{}{}.
\newblock
{\BBOQ}\APACrefatitle {Adaptive {{Rejection Metropolis Sampling}} within
  {{Gibbs Sampling}}} {Adaptive {{Rejection Metropolis Sampling}} within
  {{Gibbs Sampling}}}.{\BBCQ}
\newblock
\APACjournalVolNumPages{Applied Statistics}{44}{4}{455}.
\newblock
\begin{APACrefDOI} \doi{10.2307/2986138} \end{APACrefDOI}
\PrintBackRefs{\CurrentBib}

\bibitem [\protect \citeauthoryear {%
Gilks%
, Richardson%
\BCBL {}\ \BBA {} Spiegelhalter%
}{%
Gilks%
\ \protect \BOthers {.}}{%
{\protect \APACyear {1996}}%
}]{%
gilksMarkovChainMonte1996}
\APACinsertmetastar {%
gilksMarkovChainMonte1996}%
\begin{APACrefauthors}%
Gilks, W\BPBI R.%
, Richardson, S.%
\BCBL {}\ \BBA {} Spiegelhalter, D.%
\end{APACrefauthors}%
\ (\BEDS).
\unskip\
\newblock
\APACrefYear{1996}.
\newblock
\APACrefbtitle {Markov {{Chain Monte Carlo}} in {{Practice}}} {Markov {{Chain
  Monte Carlo}} in {{Practice}}}\ (\PrintOrdinal{1st edition}\ \BEd).
\newblock
\APACaddressPublisher{{Boca Raton, Fla}}{{Chapman and Hall/CRC}}.
\PrintBackRefs{\CurrentBib}

\bibitem [\protect \citeauthoryear {%
Green%
}{%
Green%
}{%
{\protect \APACyear {1995}}%
}]{%
greenReversibleJumpMarkov1995a}
\APACinsertmetastar {%
greenReversibleJumpMarkov1995a}%
\begin{APACrefauthors}%
Green, P\BPBI J.%
\end{APACrefauthors}%
\unskip\
\newblock
\APACrefYearMonthDay{1995}{{\APACmonth{12}}}{}.
\newblock
{\BBOQ}\APACrefatitle {Reversible Jump {{Markov}} Chain {{Monte Carlo}}
  Computation and {{Bayesian}} Model Determination} {Reversible jump {{Markov}}
  chain {{Monte Carlo}} computation and {{Bayesian}} model
  determination}.{\BBCQ}
\newblock
\APACjournalVolNumPages{Biometrika}{82}{4}{711--732}.
\newblock
\begin{APACrefDOI} \doi{10.1093/biomet/82.4.711} \end{APACrefDOI}
\PrintBackRefs{\CurrentBib}

\bibitem [\protect \citeauthoryear {%
Griewank%
\ \BBA {} Walther%
}{%
Griewank%
\ \BBA {} Walther%
}{%
{\protect \APACyear {2008}}%
}]{%
griewankEvaluatingDerivativesPrinciples2008}
\APACinsertmetastar {%
griewankEvaluatingDerivativesPrinciples2008}%
\begin{APACrefauthors}%
Griewank, A.%
\BCBT {}\ \BBA {} Walther, A.%
\end{APACrefauthors}%
\unskip\
\newblock
\APACrefYear{2008}.
\newblock
\APACrefbtitle {Evaluating Derivatives: Principles and Techniques of
  Algorithmic Differentiation} {Evaluating derivatives: Principles and
  techniques of algorithmic differentiation}\ (\PrintOrdinal{2nd ed}\ \BEd).
\newblock
\APACaddressPublisher{{Philadelphia, PA}}{{Society for Industrial and Applied
  Mathematics}}.
\PrintBackRefs{\CurrentBib}

\bibitem [\protect \citeauthoryear {%
Hansen%
, Cordua%
\BCBL {}\ \BBA {} Mosegaard%
}{%
Hansen%
\ \protect \BOthers {.}}{%
{\protect \APACyear {2012}}%
}]{%
hansenInverseProblemsNontrivial2012a}
\APACinsertmetastar {%
hansenInverseProblemsNontrivial2012a}%
\begin{APACrefauthors}%
Hansen, T\BPBI M.%
, Cordua, K\BPBI S.%
\BCBL {}\ \BBA {} Mosegaard, K.%
\end{APACrefauthors}%
\unskip\
\newblock
\APACrefYearMonthDay{2012}{{\APACmonth{06}}}{}.
\newblock
{\BBOQ}\APACrefatitle {Inverse Problems with Non-Trivial Priors: Efficient
  Solution through Sequential {{Gibbs}} Sampling} {Inverse problems with
  non-trivial priors: Efficient solution through sequential {{Gibbs}}
  sampling}.{\BBCQ}
\newblock
\APACjournalVolNumPages{Computational Geosciences}{16}{3}{593--611}.
\newblock
\begin{APACrefDOI} \doi{10.1007/s10596-011-9271-1} \end{APACrefDOI}
\PrintBackRefs{\CurrentBib}

\bibitem [\protect \citeauthoryear {%
Hansen%
, Cordua%
, Zunino%
\BCBL {}\ \BBA {} Mosegaard%
}{%
Hansen%
\ \protect \BOthers {.}}{%
{\protect \APACyear {2016}}%
}]{%
hansenProbabilisticIntegrationGeoInformation2016a}
\APACinsertmetastar {%
hansenProbabilisticIntegrationGeoInformation2016a}%
\begin{APACrefauthors}%
Hansen, T\BPBI M.%
, Cordua, K\BPBI S.%
, Zunino, A.%
\BCBL {}\ \BBA {} Mosegaard, K.%
\end{APACrefauthors}%
\unskip\
\newblock
\APACrefYearMonthDay{2016}{}{}.
\newblock
{\BBOQ}\APACrefatitle {Probabilistic {{Integration}} of {{Geo-Information}}}
  {Probabilistic {{Integration}} of {{Geo-Information}}}.{\BBCQ}
\newblock
\BIn{} \APACrefbtitle {Integrated {{Imaging}} of the {{Earth}}} {Integrated
  {{Imaging}} of the {{Earth}}}\ (\BPGS\ 93--116).
\newblock
\APACaddressPublisher{}{{American Geophysical Union (AGU)}}.
\newblock
\begin{APACrefDOI} \doi{10.1002/9781118929063.ch6} \end{APACrefDOI}
\PrintBackRefs{\CurrentBib}

\bibitem [\protect \citeauthoryear {%
Hastings%
}{%
Hastings%
}{%
{\protect \APACyear {1970}}%
}]{%
hastingsMonteCarloSampling1970}
\APACinsertmetastar {%
hastingsMonteCarloSampling1970}%
\begin{APACrefauthors}%
Hastings, W\BPBI K.%
\end{APACrefauthors}%
\unskip\
\newblock
\APACrefYearMonthDay{1970}{}{}.
\newblock
{\BBOQ}\APACrefatitle {Monte {{Carlo Sampling Methods Using Markov Chains}} and
  {{Their Applications}}} {Monte {{Carlo Sampling Methods Using Markov Chains}}
  and {{Their Applications}}}.{\BBCQ}
\newblock
\APACjournalVolNumPages{Biometrika}{57}{1}{97--109}.
\PrintBackRefs{\CurrentBib}

\bibitem [\protect \citeauthoryear {%
Hinze%
, Pinnau%
, Ulbrich%
\BCBL {}\ \BBA {} Ulbrich%
}{%
Hinze%
\ \protect \BOthers {.}}{%
{\protect \APACyear {2008}}%
}]{%
hinzeOptimizationPDEConstraints2008}
\APACinsertmetastar {%
hinzeOptimizationPDEConstraints2008}%
\begin{APACrefauthors}%
Hinze, M.%
, Pinnau, R.%
, Ulbrich, M.%
\BCBL {}\ \BBA {} Ulbrich, S.%
\end{APACrefauthors}%
\unskip\
\newblock
\APACrefYear{2008}.
\newblock
\APACrefbtitle {Optimization with {{PDE Constraints}}} {Optimization with {{PDE
  Constraints}}}\ (\PrintOrdinal{2009th edition}\ \BEd).
\newblock
\APACaddressPublisher{{New York}}{{Springer}}.
\PrintBackRefs{\CurrentBib}

\bibitem [\protect \citeauthoryear {%
Hoffman%
\ \BBA {} Gelman%
}{%
Hoffman%
\ \BBA {} Gelman%
}{%
{\protect \APACyear {2014}}%
}]{%
hoffmanNoUturnSamplerAdaptively2014}
\APACinsertmetastar {%
hoffmanNoUturnSamplerAdaptively2014}%
\begin{APACrefauthors}%
Hoffman, M\BPBI D.%
\BCBT {}\ \BBA {} Gelman, A.%
\end{APACrefauthors}%
\unskip\
\newblock
\APACrefYearMonthDay{2014}{}{}.
\newblock
{\BBOQ}\APACrefatitle {The {{No-U-turn}} Sampler: Adaptively Setting Path
  Lengths in {{Hamiltonian Monte Carlo}}} {The {{No-U-turn}} sampler:
  Adaptively setting path lengths in {{Hamiltonian Monte Carlo}}}.{\BBCQ}
\newblock
\APACjournalVolNumPages{The Journal of Machine Learning
  Research}{15}{1}{1593--1623}.
\PrintBackRefs{\CurrentBib}

\bibitem [\protect \citeauthoryear {%
{Keilis-Borok}%
\ \BBA {} Yanovskaja%
}{%
{Keilis-Borok}%
\ \BBA {} Yanovskaja%
}{%
{\protect \APACyear {1967}}%
}]{%
keilis-borokInverseProblemsSeismology1967}
\APACinsertmetastar {%
keilis-borokInverseProblemsSeismology1967}%
\begin{APACrefauthors}%
{Keilis-Borok}, V\BPBI I.%
\BCBT {}\ \BBA {} Yanovskaja, T\BPBI B.%
\end{APACrefauthors}%
\unskip\
\newblock
\APACrefYearMonthDay{1967}{}{}.
\newblock
{\BBOQ}\APACrefatitle {Inverse {{Problems}} of {{Seismology}} ({{Structural
  Review}})} {Inverse {{Problems}} of {{Seismology}} ({{Structural
  Review}})}.{\BBCQ}
\newblock
\APACjournalVolNumPages{Geophysical Journal of the Royal Astronomical
  Society}{13}{1-3}{223--234}.
\newblock
\begin{APACrefDOI} \doi{10.1111/j.1365-246X.1967.tb02156.x} \end{APACrefDOI}
\PrintBackRefs{\CurrentBib}

\bibitem [\protect \citeauthoryear {%
Khoshkholgh%
, Zunino%
\BCBL {}\ \BBA {} Mosegaard%
}{%
Khoshkholgh%
\ \protect \BOthers {.}}{%
{\protect \APACyear {2021}}%
}]{%
khoshkholghInformedProposalMonte2021}
\APACinsertmetastar {%
khoshkholghInformedProposalMonte2021}%
\begin{APACrefauthors}%
Khoshkholgh, S.%
, Zunino, A.%
\BCBL {}\ \BBA {} Mosegaard, K.%
\end{APACrefauthors}%
\unskip\
\newblock
\APACrefYearMonthDay{2021}{{\APACmonth{08}}}{}.
\newblock
{\BBOQ}\APACrefatitle {Informed Proposal {{Monte Carlo}}} {Informed proposal
  {{Monte Carlo}}}.{\BBCQ}
\newblock
\APACjournalVolNumPages{Geophysical Journal International}{226}{2}{1239--1248}.
\newblock
\begin{APACrefDOI} \doi{10.1093/gji/ggab173} \end{APACrefDOI}
\PrintBackRefs{\CurrentBib}

\bibitem [\protect \citeauthoryear {%
Khoshkholgh%
, Zunino%
\BCBL {}\ \BBA {} Mosegaard%
}{%
Khoshkholgh%
\ \protect \BOthers {.}}{%
{\protect \APACyear {2022}}%
}]{%
khoshkholghFullwaveformInversionInformedproposal2022}
\APACinsertmetastar {%
khoshkholghFullwaveformInversionInformedproposal2022}%
\begin{APACrefauthors}%
Khoshkholgh, S.%
, Zunino, A.%
\BCBL {}\ \BBA {} Mosegaard, K.%
\end{APACrefauthors}%
\unskip\
\newblock
\APACrefYearMonthDay{2022}{{\APACmonth{09}}}{}.
\newblock
{\BBOQ}\APACrefatitle {Full-Waveform Inversion by Informed-Proposal {{Monte
  Carlo}}} {Full-waveform inversion by informed-proposal {{Monte
  Carlo}}}.{\BBCQ}
\newblock
\APACjournalVolNumPages{Geophysical Journal International}{230}{3}{1824--1833}.
\newblock
\begin{APACrefDOI} \doi{10.1093/gji/ggac150} \end{APACrefDOI}
\PrintBackRefs{\CurrentBib}

\bibitem [\protect \citeauthoryear {%
Klaasen%
, Thrastarson%
, Fichtner%
, {\c C}ubuk-Sabuncu%
\BCBL {}\ \BBA {} J{\'o}nsd{\'o}ttir%
}{%
Klaasen%
\ \protect \BOthers {.}}{%
{\protect \APACyear {2022}}%
}]{%
klaasenSensingIcelandMost2022}
\APACinsertmetastar {%
klaasenSensingIcelandMost2022}%
\begin{APACrefauthors}%
Klaasen, S.%
, Thrastarson, S.%
, Fichtner, A.%
, {\c C}ubuk-Sabuncu, Y.%
\BCBL {}\ \BBA {} J{\'o}nsd{\'o}ttir, K.%
\end{APACrefauthors}%
\unskip\
\newblock
\APACrefYearMonthDay{2022}{{\APACmonth{01}}}{}.
\newblock
{\BBOQ}\APACrefatitle {Sensing {{Iceland}}'s {{Most Active Volcano}} with a
  ``{{Buried Hair}}''} {Sensing {{Iceland}}'s {{Most Active Volcano}} with a
  ``{{Buried Hair}}''}.{\BBCQ}
\newblock
\APACjournalVolNumPages{Eos}{103}{}{}.
\newblock
\begin{APACrefDOI} \doi{10.1029/2022EO220007} \end{APACrefDOI}
\PrintBackRefs{\CurrentBib}

\bibitem [\protect \citeauthoryear {%
Komatitsch%
\ \BBA {} Martin%
}{%
Komatitsch%
\ \BBA {} Martin%
}{%
{\protect \APACyear {2007}}%
}]{%
komatitschUnsplitConvolutionalPerfectly2007}
\APACinsertmetastar {%
komatitschUnsplitConvolutionalPerfectly2007}%
\begin{APACrefauthors}%
Komatitsch, D.%
\BCBT {}\ \BBA {} Martin, R.%
\end{APACrefauthors}%
\unskip\
\newblock
\APACrefYearMonthDay{2007}{{\APACmonth{09}}}{}.
\newblock
{\BBOQ}\APACrefatitle {An Unsplit Convolutional Perfectly Matched Layer
  Improved at Grazing Incidence for the Seismic Wave Equation} {An unsplit
  convolutional perfectly matched layer improved at grazing incidence for the
  seismic wave equation}.{\BBCQ}
\newblock
\APACjournalVolNumPages{Geophysics}{72}{5}{SM155-SM167}.
\newblock
\begin{APACrefDOI} \doi{10.1190/1.2757586} \end{APACrefDOI}
\PrintBackRefs{\CurrentBib}

\bibitem [\protect \citeauthoryear {%
Leung%
\ \BBA {} Qian%
}{%
Leung%
\ \BBA {} Qian%
}{%
{\protect \APACyear {2006}}%
}]{%
leungAdjointStateMethod2006}
\APACinsertmetastar {%
leungAdjointStateMethod2006}%
\begin{APACrefauthors}%
Leung, S.%
\BCBT {}\ \BBA {} Qian, J.%
\end{APACrefauthors}%
\unskip\
\newblock
\APACrefYearMonthDay{2006}{}{}.
\newblock
{\BBOQ}\APACrefatitle {An {{Adjoint State Method For Three-dimensional
  Transmission Traveltime Tomography Using First-Arrivals}}} {An {{Adjoint
  State Method For Three-dimensional Transmission Traveltime Tomography Using
  First-Arrivals}}}.{\BBCQ}
\newblock
\APACjournalVolNumPages{Communications in Mathematical
  Sciences}{4}{1}{249--266}.
\newblock
\begin{APACrefDOI} \doi{https://dx.doi.org/10.4310/CMS.2006.v4.n1.a10}
  \end{APACrefDOI}
\PrintBackRefs{\CurrentBib}

\bibitem [\protect \citeauthoryear {%
Lions%
}{%
Lions%
}{%
{\protect \APACyear {1971}}%
}]{%
lionsOptimalControlSystems1971}
\APACinsertmetastar {%
lionsOptimalControlSystems1971}%
\begin{APACrefauthors}%
Lions, J\BHBI L.%
\end{APACrefauthors}%
\unskip\
\newblock
\APACrefYear{1971}.
\newblock
\APACrefbtitle {Optimal {{Control}} of {{Systems Governed}} by {{Partial
  Differential Equations}}} {Optimal {{Control}} of {{Systems Governed}} by
  {{Partial Differential Equations}}}.
\newblock
\APACaddressPublisher{}{{Springer-Verlag}}.
\PrintBackRefs{\CurrentBib}

\bibitem [\protect \citeauthoryear {%
J\BPBI S.~Liu%
\ \BBA {} Chen%
}{%
J\BPBI S.~Liu%
\ \BBA {} Chen%
}{%
{\protect \APACyear {1998}}%
}]{%
liuSequentialMonteCarlo1998}
\APACinsertmetastar {%
liuSequentialMonteCarlo1998}%
\begin{APACrefauthors}%
Liu, J\BPBI S.%
\BCBT {}\ \BBA {} Chen, R.%
\end{APACrefauthors}%
\unskip\
\newblock
\APACrefYearMonthDay{1998}{{\APACmonth{09}}}{}.
\newblock
{\BBOQ}\APACrefatitle {Sequential {{Monte Carlo Methods}} for {{Dynamic
  Systems}}} {Sequential {{Monte Carlo Methods}} for {{Dynamic
  Systems}}}.{\BBCQ}
\newblock
\APACjournalVolNumPages{Journal of the American Statistical
  Association}{93}{443}{1032--1044}.
\newblock
\begin{APACrefDOI} \doi{10.1080/01621459.1998.10473765} \end{APACrefDOI}
\PrintBackRefs{\CurrentBib}

\bibitem [\protect \citeauthoryear {%
Q.~Liu%
\ \BBA {} Wang%
}{%
Q.~Liu%
\ \BBA {} Wang%
}{%
{\protect \APACyear {2019}}%
}]{%
liuSteinVariationalGradient2019}
\APACinsertmetastar {%
liuSteinVariationalGradient2019}%
\begin{APACrefauthors}%
Liu, Q.%
\BCBT {}\ \BBA {} Wang, D.%
\end{APACrefauthors}%
\unskip\
\newblock
\APACrefYearMonthDay{2019}{{\APACmonth{09}}}{}.
\newblock
\APACrefbtitle {Stein {{Variational Gradient Descent}}: {{A General Purpose
  Bayesian Inference Algorithm}}} {Stein {{Variational Gradient Descent}}: {{A
  General Purpose Bayesian Inference Algorithm}}}\ (\BNUM\ arXiv:1608.04471).
\newblock
\APACaddressPublisher{}{{arXiv}}.
\newblock
\begin{APACrefDOI} \doi{10.48550/arXiv.1608.04471} \end{APACrefDOI}
\PrintBackRefs{\CurrentBib}

\bibitem [\protect \citeauthoryear {%
Malinverno%
}{%
Malinverno%
}{%
{\protect \APACyear {2002}}%
}]{%
malinvernoParsimoniousBayesianMarkov2002}
\APACinsertmetastar {%
malinvernoParsimoniousBayesianMarkov2002}%
\begin{APACrefauthors}%
Malinverno, A.%
\end{APACrefauthors}%
\unskip\
\newblock
\APACrefYearMonthDay{2002}{}{}.
\newblock
{\BBOQ}\APACrefatitle {Parsimonious {{Bayesian Markov}} Chain {{Monte Carlo}}
  Inversion in a Nonlinear Geophysical Problem} {Parsimonious {{Bayesian
  Markov}} chain {{Monte Carlo}} inversion in a nonlinear geophysical
  problem}.{\BBCQ}
\newblock
\APACjournalVolNumPages{Geophysical Journal International}{151}{3}{675--688}.
\newblock
\begin{APACrefDOI} \doi{10.1046/j.1365-246X.2002.01847.x} \end{APACrefDOI}
\PrintBackRefs{\CurrentBib}

\bibitem [\protect \citeauthoryear {%
Mavko%
, Mukerji%
\BCBL {}\ \BBA {} Dvorkin%
}{%
Mavko%
\ \protect \BOthers {.}}{%
{\protect \APACyear {2003}}%
}]{%
mavkoRockPhysicsHandbook2003a}
\APACinsertmetastar {%
mavkoRockPhysicsHandbook2003a}%
\begin{APACrefauthors}%
Mavko, G.%
, Mukerji, T.%
\BCBL {}\ \BBA {} Dvorkin, J.%
\end{APACrefauthors}%
\unskip\
\newblock
\APACrefYear{2003}.
\newblock
\APACrefbtitle {The {{Rock Physics Handbook}}: {{Tools}} for {{Seismic
  Analysis}} of {{Porous Media}}} {The {{Rock Physics Handbook}}: {{Tools}} for
  {{Seismic Analysis}} of {{Porous Media}}}.
\newblock
\begin{APACrefDOI} \doi{10.1017/CBO9780511626753} \end{APACrefDOI}
\PrintBackRefs{\CurrentBib}

\bibitem [\protect \citeauthoryear {%
Metropolis%
, Rosenbluth%
, Rosenbluth%
, Teller%
\BCBL {}\ \BBA {} Teller%
}{%
Metropolis%
\ \protect \BOthers {.}}{%
{\protect \APACyear {1953}}%
}]{%
metropolisEquationStateCalculations1953a}
\APACinsertmetastar {%
metropolisEquationStateCalculations1953a}%
\begin{APACrefauthors}%
Metropolis, N.%
, Rosenbluth, A\BPBI W.%
, Rosenbluth, M\BPBI N.%
, Teller, A\BPBI H.%
\BCBL {}\ \BBA {} Teller, E.%
\end{APACrefauthors}%
\unskip\
\newblock
\APACrefYearMonthDay{1953}{}{}.
\newblock
{\BBOQ}\APACrefatitle {Equation of {{State Calculations}} by {{Fast Computing
  Machines}}} {Equation of {{State Calculations}} by {{Fast Computing
  Machines}}}.{\BBCQ}
\newblock
\APACjournalVolNumPages{The Journal of Chemical Physics}{21}{6}{1087--1092}.
\newblock
\begin{APACrefDOI} \doi{10.1063/1.1699114} \end{APACrefDOI}
\PrintBackRefs{\CurrentBib}

\bibitem [\protect \citeauthoryear {%
Metropolis%
\ \BBA {} Ulam%
}{%
Metropolis%
\ \BBA {} Ulam%
}{%
{\protect \APACyear {1949}}%
}]{%
metropolisMonteCarloMethod1949}
\APACinsertmetastar {%
metropolisMonteCarloMethod1949}%
\begin{APACrefauthors}%
Metropolis, N.%
\BCBT {}\ \BBA {} Ulam, S.%
\end{APACrefauthors}%
\unskip\
\newblock
\APACrefYearMonthDay{1949}{}{}.
\newblock
{\BBOQ}\APACrefatitle {The {{Monte Carlo}} Method} {The {{Monte Carlo}}
  method}.{\BBCQ}
\newblock
\APACjournalVolNumPages{Journal of the American Statistical
  Association}{44}{247}{335--341}.
\PrintBackRefs{\CurrentBib}

\bibitem [\protect \citeauthoryear {%
Mosegaard%
}{%
Mosegaard%
}{%
{\protect \APACyear {2011}}%
}]{%
mosegaardQuestConsistencySymmetry2011}
\APACinsertmetastar {%
mosegaardQuestConsistencySymmetry2011}%
\begin{APACrefauthors}%
Mosegaard, K.%
\end{APACrefauthors}%
\unskip\
\newblock
\APACrefYearMonthDay{2011}{{\APACmonth{09}}}{}.
\newblock
{\BBOQ}\APACrefatitle {Quest for Consistency, Symmetry, and Simplicity
  \textemdash{} {{The}} Legacy of {{Albert Tarantola}}} {Quest for consistency,
  symmetry, and simplicity \textemdash{} {{The}} legacy of {{Albert
  Tarantola}}}.{\BBCQ}
\newblock
\APACjournalVolNumPages{Geophysics}{76}{5}{W51-W61}.
\newblock
\begin{APACrefDOI} \doi{10.1190/geo2010-0328.1} \end{APACrefDOI}
\PrintBackRefs{\CurrentBib}

\bibitem [\protect \citeauthoryear {%
Mosegaard%
\ \BBA {} Tarantola%
}{%
Mosegaard%
\ \BBA {} Tarantola%
}{%
{\protect \APACyear {1995}}%
}]{%
mosegaardMonteCarloSampling1995}
\APACinsertmetastar {%
mosegaardMonteCarloSampling1995}%
\begin{APACrefauthors}%
Mosegaard, K.%
\BCBT {}\ \BBA {} Tarantola, A.%
\end{APACrefauthors}%
\unskip\
\newblock
\APACrefYearMonthDay{1995}{}{}.
\newblock
{\BBOQ}\APACrefatitle {Monte {{Carlo}} Sampling of Solutions to Inverse
  Problems} {Monte {{Carlo}} sampling of solutions to inverse problems}.{\BBCQ}
\newblock
\APACjournalVolNumPages{Journal of Geophysical Research: Solid
  Earth}{100}{B7}{12431--12447}.
\newblock
\begin{APACrefDOI} \doi{10.1029/94JB03097} \end{APACrefDOI}
\PrintBackRefs{\CurrentBib}

\bibitem [\protect \citeauthoryear {%
Mosegaard%
\ \BBA {} Tarantola%
}{%
Mosegaard%
\ \BBA {} Tarantola%
}{%
{\protect \APACyear {2002}}%
}]{%
mosegaardProbabilisticApproachInverse2002}
\APACinsertmetastar {%
mosegaardProbabilisticApproachInverse2002}%
\begin{APACrefauthors}%
Mosegaard, K.%
\BCBT {}\ \BBA {} Tarantola, A.%
\end{APACrefauthors}%
\unskip\
\newblock
\APACrefYearMonthDay{2002}{}{}.
\newblock
{\BBOQ}\APACrefatitle {Probabilistic {{Approach}} to {{Inverse Problems}}}
  {Probabilistic {{Approach}} to {{Inverse Problems}}}.{\BBCQ}
\newblock
\BIn{} \APACrefbtitle {In the {{International Handbook}} of {{Earthquake}} \&
  {{Engineering Seismology}} ({{Part A}}} {In the {{International Handbook}} of
  {{Earthquake}} \& {{Engineering Seismology}} ({{Part A}}}\ (\BPGS\ 237--265).
\newblock
\APACaddressPublisher{}{{Academic Press}}.
\PrintBackRefs{\CurrentBib}

\bibitem [\protect \citeauthoryear {%
Neal%
}{%
Neal%
}{%
{\protect \APACyear {2003}}%
}]{%
nealSliceSampling2003}
\APACinsertmetastar {%
nealSliceSampling2003}%
\begin{APACrefauthors}%
Neal, R.%
\end{APACrefauthors}%
\unskip\
\newblock
\APACrefYearMonthDay{2003}{{\APACmonth{06}}}{}.
\newblock
{\BBOQ}\APACrefatitle {Slice Sampling} {Slice sampling}.{\BBCQ}
\newblock
\APACjournalVolNumPages{The Annals of Statistics}{31}{3}{705--767}.
\newblock
\begin{APACrefDOI} \doi{10.1214/aos/1056562461} \end{APACrefDOI}
\PrintBackRefs{\CurrentBib}

\bibitem [\protect \citeauthoryear {%
Neal%
}{%
Neal%
}{%
{\protect \APACyear {2012}}%
}]{%
nealMCMCUsingHamiltonian2012}
\APACinsertmetastar {%
nealMCMCUsingHamiltonian2012}%
\begin{APACrefauthors}%
Neal, R.%
\end{APACrefauthors}%
\unskip\
\newblock
\APACrefYearMonthDay{2012}{}{}.
\newblock
{\BBOQ}\APACrefatitle {{{MCMC}} Using {{Hamiltonian}} Dynamics} {{{MCMC}} using
  {{Hamiltonian}} dynamics}.{\BBCQ}
\newblock
\APACjournalVolNumPages{Handbook of Markov Chain Monte Carlo}{}{}{}.
\newblock
\begin{APACrefDOI} \doi{10.1201/b10905-6} \end{APACrefDOI}
\PrintBackRefs{\CurrentBib}

\bibitem [\protect \citeauthoryear {%
Nocedal%
\ \BBA {} Wright%
}{%
Nocedal%
\ \BBA {} Wright%
}{%
{\protect \APACyear {2006}}%
}]{%
nocedalNumericalOptimization2006}
\APACinsertmetastar {%
nocedalNumericalOptimization2006}%
\begin{APACrefauthors}%
Nocedal, J.%
\BCBT {}\ \BBA {} Wright, S\BPBI J.%
\end{APACrefauthors}%
\unskip\
\newblock
\APACrefYear{2006}.
\newblock
\APACrefbtitle {Numerical Optimization} {Numerical optimization}\
  (\PrintOrdinal{2nd ed}\ \BEd).
\newblock
\APACaddressPublisher{{New York}}{{Springer}}.
\PrintBackRefs{\CurrentBib}

\bibitem [\protect \citeauthoryear {%
Pasalic%
\ \BBA {} McGarry%
}{%
Pasalic%
\ \BBA {} McGarry%
}{%
{\protect \APACyear {2010}}%
}]{%
pasalicConvolutionalPerfectlyMatched2010}
\APACinsertmetastar {%
pasalicConvolutionalPerfectlyMatched2010}%
\begin{APACrefauthors}%
Pasalic, D.%
\BCBT {}\ \BBA {} McGarry, R.%
\end{APACrefauthors}%
\unskip\
\newblock
\APACrefYearMonthDay{2010}{{\APACmonth{01}}}{}.
\newblock
{\BBOQ}\APACrefatitle {Convolutional Perfectly Matched Layer for Isotropic and
  Anisotropic Acoustic Wave Equations} {Convolutional perfectly matched layer
  for isotropic and anisotropic acoustic wave equations}.{\BBCQ}
\newblock
\BIn{} \APACrefbtitle {{{SEG Technical Program Expanded Abstracts}} 2010}
  {{{SEG Technical Program Expanded Abstracts}} 2010}\ (\BPGS\ 2925--2929).
\newblock
\APACaddressPublisher{}{{Society of Exploration Geophysicists}}.
\newblock
\begin{APACrefDOI} \doi{10.1190/1.3513453} \end{APACrefDOI}
\PrintBackRefs{\CurrentBib}

\bibitem [\protect \citeauthoryear {%
Plessix%
}{%
Plessix%
}{%
{\protect \APACyear {2006}}%
}]{%
plessixReviewAdjointstateMethod2006}
\APACinsertmetastar {%
plessixReviewAdjointstateMethod2006}%
\begin{APACrefauthors}%
Plessix, R\BHBI E.%
\end{APACrefauthors}%
\unskip\
\newblock
\APACrefYearMonthDay{2006}{}{}.
\newblock
{\BBOQ}\APACrefatitle {A Review of the Adjoint-State Method for Computing the
  Gradient of a Functional with Geophysical Applications} {A review of the
  adjoint-state method for computing the gradient of a functional with
  geophysical applications}.{\BBCQ}
\newblock
\APACjournalVolNumPages{Geophysical Journal International}{167}{2}{495--503}.
\newblock
\begin{APACrefDOI} \doi{10.1111/j.1365-246X.2006.02978.x} \end{APACrefDOI}
\PrintBackRefs{\CurrentBib}

\bibitem [\protect \citeauthoryear {%
Press%
}{%
Press%
}{%
{\protect \APACyear {1968}}%
}]{%
pressEarthModelsObtained1968}
\APACinsertmetastar {%
pressEarthModelsObtained1968}%
\begin{APACrefauthors}%
Press, F.%
\end{APACrefauthors}%
\unskip\
\newblock
\APACrefYearMonthDay{1968}{}{}.
\newblock
{\BBOQ}\APACrefatitle {Earth Models Obtained by {{Monte Carlo Inversion}}}
  {Earth models obtained by {{Monte Carlo Inversion}}}.{\BBCQ}
\newblock
\APACjournalVolNumPages{Journal of Geophysical Research
  (1896-1977)}{73}{16}{5223--5234}.
\newblock
\begin{APACrefDOI} \doi{10.1029/JB073i016p05223} \end{APACrefDOI}
\PrintBackRefs{\CurrentBib}

\bibitem [\protect \citeauthoryear {%
Rasmussen%
\ \BBA {} Pedersen%
}{%
Rasmussen%
\ \BBA {} Pedersen%
}{%
{\protect \APACyear {1979}}%
}]{%
rasmussenEndCorrectionsPotential1979a}
\APACinsertmetastar {%
rasmussenEndCorrectionsPotential1979a}%
\begin{APACrefauthors}%
Rasmussen, R.%
\BCBT {}\ \BBA {} Pedersen, L\BPBI B.%
\end{APACrefauthors}%
\unskip\
\newblock
\APACrefYearMonthDay{1979}{}{}.
\newblock
{\BBOQ}\APACrefatitle {End {{Corrections}} in {{Potential Field Modeling}}}
  {End {{Corrections}} in {{Potential Field Modeling}}}.{\BBCQ}
\newblock
\APACjournalVolNumPages{Geophysical Prospecting}{27}{4}{749--760}.
\newblock
\begin{APACrefDOI} \doi{10.1111/j.1365-2478.1979.tb00994.x} \end{APACrefDOI}
\PrintBackRefs{\CurrentBib}

\bibitem [\protect \citeauthoryear {%
Rawlinson%
\ \BBA {} Sambridge%
}{%
Rawlinson%
\ \BBA {} Sambridge%
}{%
{\protect \APACyear {2004}}%
}]{%
rawlinsonWaveFrontEvolution2004}
\APACinsertmetastar {%
rawlinsonWaveFrontEvolution2004}%
\begin{APACrefauthors}%
Rawlinson, N.%
\BCBT {}\ \BBA {} Sambridge, M.%
\end{APACrefauthors}%
\unskip\
\newblock
\APACrefYearMonthDay{2004}{{\APACmonth{03}}}{}.
\newblock
{\BBOQ}\APACrefatitle {Wave Front Evolution in Strongly Heterogeneous Layered
  Media Using the Fast Marching Method} {Wave front evolution in strongly
  heterogeneous layered media using the fast marching method}.{\BBCQ}
\newblock
\APACjournalVolNumPages{Geophysical Journal International}{156}{3}{631--647}.
\newblock
\begin{APACrefDOI} \doi{10.1111/j.1365-246X.2004.02153.x} \end{APACrefDOI}
\PrintBackRefs{\CurrentBib}

\bibitem [\protect \citeauthoryear {%
Sambridge%
}{%
Sambridge%
}{%
{\protect \APACyear {1999}}%
}]{%
sambridgeGeophysicalInversionNeighbourhood1999}
\APACinsertmetastar {%
sambridgeGeophysicalInversionNeighbourhood1999}%
\begin{APACrefauthors}%
Sambridge, M.%
\end{APACrefauthors}%
\unskip\
\newblock
\APACrefYearMonthDay{1999}{{\APACmonth{08}}}{}.
\newblock
{\BBOQ}\APACrefatitle {Geophysical Inversion with a Neighbourhood
  Algorithm\textemdash{{I}}. {{Searching}} a Parameter Space} {Geophysical
  inversion with a neighbourhood algorithm\textemdash{{I}}. {{Searching}} a
  parameter space}.{\BBCQ}
\newblock
\APACjournalVolNumPages{Geophysical Journal International}{138}{2}{479--494}.
\newblock
\begin{APACrefDOI} \doi{10.1046/j.1365-246X.1999.00876.x} \end{APACrefDOI}
\PrintBackRefs{\CurrentBib}

\bibitem [\protect \citeauthoryear {%
Sambridge%
}{%
Sambridge%
}{%
{\protect \APACyear {2014}}%
}]{%
sambridgeParallelTemperingAlgorithm2014}
\APACinsertmetastar {%
sambridgeParallelTemperingAlgorithm2014}%
\begin{APACrefauthors}%
Sambridge, M.%
\end{APACrefauthors}%
\unskip\
\newblock
\APACrefYearMonthDay{2014}{{\APACmonth{01}}}{}.
\newblock
{\BBOQ}\APACrefatitle {A {{Parallel Tempering}} Algorithm for Probabilistic
  Sampling and Multimodal Optimization} {A {{Parallel Tempering}} algorithm for
  probabilistic sampling and multimodal optimization}.{\BBCQ}
\newblock
\APACjournalVolNumPages{Geophysical Journal International}{196}{1}{357--374}.
\newblock
\begin{APACrefDOI} \doi{10.1093/gji/ggt342} \end{APACrefDOI}
\PrintBackRefs{\CurrentBib}

\bibitem [\protect \citeauthoryear {%
Sambridge%
\ \BBA {} Mosegaard%
}{%
Sambridge%
\ \BBA {} Mosegaard%
}{%
{\protect \APACyear {2002}}%
}]{%
sambridgeMonteCarloMethods2002}
\APACinsertmetastar {%
sambridgeMonteCarloMethods2002}%
\begin{APACrefauthors}%
Sambridge, M.%
\BCBT {}\ \BBA {} Mosegaard, K.%
\end{APACrefauthors}%
\unskip\
\newblock
\APACrefYearMonthDay{2002}{}{}.
\newblock
{\BBOQ}\APACrefatitle {Monte {{Carlo Methods}} in {{Geophysical Inverse
  Problems}}} {Monte {{Carlo Methods}} in {{Geophysical Inverse
  Problems}}}.{\BBCQ}
\newblock
\APACjournalVolNumPages{Reviews of Geophysics}{40}{3}{3-1-3-29}.
\newblock
\begin{APACrefDOI} \doi{10.1029/2000RG000089} \end{APACrefDOI}
\PrintBackRefs{\CurrentBib}

\bibitem [\protect \citeauthoryear {%
Sambridge%
, Rickwood%
, Rawlinson%
\BCBL {}\ \BBA {} Sommacal%
}{%
Sambridge%
\ \protect \BOthers {.}}{%
{\protect \APACyear {2007}}%
}]{%
sambridgeAutomaticDifferentiationGeophysical2007}
\APACinsertmetastar {%
sambridgeAutomaticDifferentiationGeophysical2007}%
\begin{APACrefauthors}%
Sambridge, M.%
, Rickwood, P.%
, Rawlinson, N.%
\BCBL {}\ \BBA {} Sommacal, S.%
\end{APACrefauthors}%
\unskip\
\newblock
\APACrefYearMonthDay{2007}{}{}.
\newblock
{\BBOQ}\APACrefatitle {Automatic Differentiation in Geophysical Inverse
  Problems} {Automatic differentiation in geophysical inverse problems}.{\BBCQ}
\newblock
\APACjournalVolNumPages{Geophysical Journal International}{170}{1}{1--8}.
\newblock
\begin{APACrefDOI} \doi{10.1111/j.1365-246X.2007.03400.x} \end{APACrefDOI}
\PrintBackRefs{\CurrentBib}

\bibitem [\protect \citeauthoryear {%
Scales%
\ \BBA {} Tenorio%
}{%
Scales%
\ \BBA {} Tenorio%
}{%
{\protect \APACyear {2001}}%
}]{%
scalesPriorInformationUncertainty2001}
\APACinsertmetastar {%
scalesPriorInformationUncertainty2001}%
\begin{APACrefauthors}%
Scales, J.%
\BCBT {}\ \BBA {} Tenorio, L.%
\end{APACrefauthors}%
\unskip\
\newblock
\APACrefYearMonthDay{2001}{}{}.
\newblock
{\BBOQ}\APACrefatitle {Prior {{Information}} and {{Uncertainty}} in {{Inverse
  Problems}}} {Prior {{Information}} and {{Uncertainty}} in {{Inverse
  Problems}}}.{\BBCQ}
\newblock
\APACjournalVolNumPages{Geophysics}{66}{}{}.
\newblock
\begin{APACrefDOI} \doi{10.1190/1.1444930} \end{APACrefDOI}
\PrintBackRefs{\CurrentBib}

\bibitem [\protect \citeauthoryear {%
Sethian%
}{%
Sethian%
}{%
{\protect \APACyear {1996}}%
}]{%
sethianFastMarchingLevel1996}
\APACinsertmetastar {%
sethianFastMarchingLevel1996}%
\begin{APACrefauthors}%
Sethian, J\BPBI A.%
\end{APACrefauthors}%
\unskip\
\newblock
\APACrefYearMonthDay{1996}{{\APACmonth{02}}}{}.
\newblock
{\BBOQ}\APACrefatitle {A Fast Marching Level Set Method for Monotonically
  Advancing Fronts.} {A fast marching level set method for monotonically
  advancing fronts.}{\BBCQ}
\newblock
\APACjournalVolNumPages{Proceedings of the National Academy of
  Sciences}{93}{4}{1591--1595}.
\newblock
\begin{APACrefDOI} \doi{10.1073/pnas.93.4.1591} \end{APACrefDOI}
\PrintBackRefs{\CurrentBib}

\bibitem [\protect \citeauthoryear {%
Simo%
, Tarnow%
\BCBL {}\ \BBA {} Wong%
}{%
Simo%
\ \protect \BOthers {.}}{%
{\protect \APACyear {1992}}%
}]{%
simoExactEnergymomentumConserving1992c}
\APACinsertmetastar {%
simoExactEnergymomentumConserving1992c}%
\begin{APACrefauthors}%
Simo, J\BPBI C.%
, Tarnow, N.%
\BCBL {}\ \BBA {} Wong, K\BPBI K.%
\end{APACrefauthors}%
\unskip\
\newblock
\APACrefYearMonthDay{1992}{{\APACmonth{10}}}{}.
\newblock
{\BBOQ}\APACrefatitle {Exact Energy-Momentum Conserving Algorithms and
  Symplectic Schemes for Nonlinear Dynamics} {Exact energy-momentum conserving
  algorithms and symplectic schemes for nonlinear dynamics}.{\BBCQ}
\newblock
\APACjournalVolNumPages{Computer Methods in Applied Mechanics and
  Engineering}{100}{1}{63--116}.
\newblock
\begin{APACrefDOI} \doi{10.1016/0045-7825(92)90115-Z} \end{APACrefDOI}
\PrintBackRefs{\CurrentBib}

\bibitem [\protect \citeauthoryear {%
Stoffa%
\ \BBA {} Sen%
}{%
Stoffa%
\ \BBA {} Sen%
}{%
{\protect \APACyear {1991}}%
}]{%
stoffaNonlinearMultiparameterOptimization1991}
\APACinsertmetastar {%
stoffaNonlinearMultiparameterOptimization1991}%
\begin{APACrefauthors}%
Stoffa, P\BPBI L.%
\BCBT {}\ \BBA {} Sen, M\BPBI K.%
\end{APACrefauthors}%
\unskip\
\newblock
\APACrefYearMonthDay{1991}{{\APACmonth{11}}}{}.
\newblock
{\BBOQ}\APACrefatitle {Nonlinear Multiparameter Optimization Using Genetic
  Algorithms: {{Inversion}} of Plane-wave Seismograms} {Nonlinear
  multiparameter optimization using genetic algorithms: {{Inversion}} of
  plane-wave seismograms}.{\BBCQ}
\newblock
\APACjournalVolNumPages{Geophysics}{56}{11}{1794--1810}.
\newblock
\begin{APACrefDOI} \doi{10.1190/1.1442992} \end{APACrefDOI}
\PrintBackRefs{\CurrentBib}

\bibitem [\protect \citeauthoryear {%
Taillandier%
, Noble%
, Chauris%
\BCBL {}\ \BBA {} Calandra%
}{%
Taillandier%
\ \protect \BOthers {.}}{%
{\protect \APACyear {2009}}%
}]{%
taillandierFirstarrivalTraveltimeTomography2009}
\APACinsertmetastar {%
taillandierFirstarrivalTraveltimeTomography2009}%
\begin{APACrefauthors}%
Taillandier, C.%
, Noble, M.%
, Chauris, H.%
\BCBL {}\ \BBA {} Calandra, H.%
\end{APACrefauthors}%
\unskip\
\newblock
\APACrefYearMonthDay{2009}{{\APACmonth{11}}}{}.
\newblock
{\BBOQ}\APACrefatitle {First-Arrival Traveltime Tomography Based on the
  Adjoint-State Method} {First-arrival traveltime tomography based on the
  adjoint-state method}.{\BBCQ}
\newblock
\APACjournalVolNumPages{Geophysics}{74}{6}{WCB1-WCB10}.
\newblock
\begin{APACrefDOI} \doi{10.1190/1.3250266} \end{APACrefDOI}
\PrintBackRefs{\CurrentBib}

\bibitem [\protect \citeauthoryear {%
Talagrand%
\ \BBA {} Courtier%
}{%
Talagrand%
\ \BBA {} Courtier%
}{%
{\protect \APACyear {1987}}%
}]{%
talagrandVariationalAssimilationMeteorological1987}
\APACinsertmetastar {%
talagrandVariationalAssimilationMeteorological1987}%
\begin{APACrefauthors}%
Talagrand, O.%
\BCBT {}\ \BBA {} Courtier, P.%
\end{APACrefauthors}%
\unskip\
\newblock
\APACrefYearMonthDay{1987}{}{}.
\newblock
{\BBOQ}\APACrefatitle {Variational {{Assimilation}} of {{Meteorological
  Observations With}} the {{Adjoint Vorticity Equation}}. {{I}}: {{Theory}}}
  {Variational {{Assimilation}} of {{Meteorological Observations With}} the
  {{Adjoint Vorticity Equation}}. {{I}}: {{Theory}}}.{\BBCQ}
\newblock
\APACjournalVolNumPages{Quarterly Journal of the Royal Meteorological
  Society}{113}{478}{1311--1328}.
\newblock
\begin{APACrefDOI} \doi{10.1002/qj.49711347812} \end{APACrefDOI}
\PrintBackRefs{\CurrentBib}

\bibitem [\protect \citeauthoryear {%
Tarantola%
}{%
Tarantola%
}{%
{\protect \APACyear {1984}}%
}]{%
tarantolaInversionSeismicReflection1984}
\APACinsertmetastar {%
tarantolaInversionSeismicReflection1984}%
\begin{APACrefauthors}%
Tarantola, A.%
\end{APACrefauthors}%
\unskip\
\newblock
\APACrefYearMonthDay{1984}{{\APACmonth{08}}}{}.
\newblock
{\BBOQ}\APACrefatitle {Inversion of Seismic Reflection Data in the Acoustic
  Approximation} {Inversion of seismic reflection data in the acoustic
  approximation}.{\BBCQ}
\newblock
\APACjournalVolNumPages{Geophysics}{49}{8}{1259--1266}.
\newblock
\begin{APACrefDOI} \doi{10.1190/1.1441754} \end{APACrefDOI}
\PrintBackRefs{\CurrentBib}

\bibitem [\protect \citeauthoryear {%
Tarantola%
}{%
Tarantola%
}{%
{\protect \APACyear {2005}}%
}]{%
tarantolaInverseProblemTheory2005a}
\APACinsertmetastar {%
tarantolaInverseProblemTheory2005a}%
\begin{APACrefauthors}%
Tarantola, A.%
\end{APACrefauthors}%
\unskip\
\newblock
\APACrefYear{2005}.
\newblock
\APACrefbtitle {Inverse {{Problem Theory}} and {{Methods}} for {{Model
  Parameter Estimation}}} {Inverse {{Problem Theory}} and {{Methods}} for
  {{Model Parameter Estimation}}}.
\newblock
\APACaddressPublisher{}{{Society for Industrial and Applied Mathematics}}.
\newblock
\begin{APACrefDOI} \doi{10.1137/1.9780898717921} \end{APACrefDOI}
\PrintBackRefs{\CurrentBib}

\bibitem [\protect \citeauthoryear {%
Tarantola%
\ \BBA {} Valette%
}{%
Tarantola%
\ \BBA {} Valette%
}{%
{\protect \APACyear {1982}}%
}]{%
tarantolaInverseProblemsQuest1982a}
\APACinsertmetastar {%
tarantolaInverseProblemsQuest1982a}%
\begin{APACrefauthors}%
Tarantola, A.%
\BCBT {}\ \BBA {} Valette, B.%
\end{APACrefauthors}%
\unskip\
\newblock
\APACrefYearMonthDay{1982}{}{}.
\newblock
{\BBOQ}\APACrefatitle {Inverse Problems = Quest for Information} {Inverse
  problems = quest for information}.{\BBCQ}
\newblock
\APACjournalVolNumPages{J. Geophys.}{50}{}{159--170}.
\PrintBackRefs{\CurrentBib}

\bibitem [\protect \citeauthoryear {%
Treister%
\ \BBA {} Haber%
}{%
Treister%
\ \BBA {} Haber%
}{%
{\protect \APACyear {2016}}%
}]{%
treisterFastMarchingAlgorithm2016}
\APACinsertmetastar {%
treisterFastMarchingAlgorithm2016}%
\begin{APACrefauthors}%
Treister, E.%
\BCBT {}\ \BBA {} Haber, E.%
\end{APACrefauthors}%
\unskip\
\newblock
\APACrefYearMonthDay{2016}{{\APACmonth{11}}}{}.
\newblock
{\BBOQ}\APACrefatitle {A Fast Marching Algorithm for the Factored Eikonal
  Equation} {A fast marching algorithm for the factored eikonal
  equation}.{\BBCQ}
\newblock
\APACjournalVolNumPages{Journal of Computational Physics}{324}{}{210--225}.
\newblock
\begin{APACrefDOI} \doi{10.1016/j.jcp.2016.08.012} \end{APACrefDOI}
\PrintBackRefs{\CurrentBib}

\bibitem [\protect \citeauthoryear {%
Tromp%
, Tape%
\BCBL {}\ \BBA {} Liu%
}{%
Tromp%
\ \protect \BOthers {.}}{%
{\protect \APACyear {2005}}%
}]{%
trompSeismicTomographyAdjoint2005a}
\APACinsertmetastar {%
trompSeismicTomographyAdjoint2005a}%
\begin{APACrefauthors}%
Tromp, J.%
, Tape, C.%
\BCBL {}\ \BBA {} Liu, Q.%
\end{APACrefauthors}%
\unskip\
\newblock
\APACrefYearMonthDay{2005}{{\APACmonth{01}}}{}.
\newblock
{\BBOQ}\APACrefatitle {Seismic Tomography, Adjoint Methods, Time Reversal and
  Banana-Doughnut Kernels} {Seismic tomography, adjoint methods, time reversal
  and banana-doughnut kernels}.{\BBCQ}
\newblock
\APACjournalVolNumPages{Geophysical Journal International}{160}{1}{195--216}.
\newblock
\begin{APACrefDOI} \doi{10.1111/j.1365-246X.2004.02453.x} \end{APACrefDOI}
\PrintBackRefs{\CurrentBib}

\bibitem [\protect \citeauthoryear {%
{van Rossum}%
}{%
{van Rossum}%
}{%
{\protect \APACyear {1995}}%
}]{%
vanrossumPythonTutorial1995}
\APACinsertmetastar {%
vanrossumPythonTutorial1995}%
\begin{APACrefauthors}%
{van Rossum}, G.%
\end{APACrefauthors}%
\unskip\
\newblock
\APACrefYearMonthDay{1995}{}{}.
\newblock
\APACrefbtitle {Python Tutorial} {Python tutorial}\ \APACbVolEdTR{}{\BTR{}\
  \BNUM\ CS-R9526}.
\newblock
\APACaddressInstitution{{Amsterdam}}{{Centrum voor Wiskunde en Informatica
  (CWI)}}.
\PrintBackRefs{\CurrentBib}

\bibitem [\protect \citeauthoryear {%
Virtanen%
\ \protect \BOthers {.}}{%
Virtanen%
\ \protect \BOthers {.}}{%
{\protect \APACyear {2020}}%
}]{%
virtanenSciPyFundamentalAlgorithms2020}
\APACinsertmetastar {%
virtanenSciPyFundamentalAlgorithms2020}%
\begin{APACrefauthors}%
Virtanen, P.%
, Gommers, R.%
, Oliphant, T\BPBI E.%
, Haberland, M.%
, Reddy, T.%
, Cournapeau, D.%
\BDBL {}{van Mulbregt}, P.%
\end{APACrefauthors}%
\unskip\
\newblock
\APACrefYearMonthDay{2020}{{\APACmonth{03}}}{}.
\newblock
{\BBOQ}\APACrefatitle {{{SciPy}} 1.0: Fundamental Algorithms for Scientific
  Computing in {{Python}}} {{{SciPy}} 1.0: Fundamental algorithms for
  scientific computing in {{Python}}}.{\BBCQ}
\newblock
\APACjournalVolNumPages{Nature Methods}{17}{3}{261--272}.
\newblock
\begin{APACrefDOI} \doi{10.1038/s41592-019-0686-2} \end{APACrefDOI}
\PrintBackRefs{\CurrentBib}

\bibitem [\protect \citeauthoryear {%
Wolpert%
\ \BBA {} Macready%
}{%
Wolpert%
\ \BBA {} Macready%
}{%
{\protect \APACyear {1997}}%
}]{%
wolpertNoFreeLunch1997}
\APACinsertmetastar {%
wolpertNoFreeLunch1997}%
\begin{APACrefauthors}%
Wolpert, D.%
\BCBT {}\ \BBA {} Macready, W.%
\end{APACrefauthors}%
\unskip\
\newblock
\APACrefYearMonthDay{1997}{{\APACmonth{04}}}{}.
\newblock
{\BBOQ}\APACrefatitle {No Free Lunch Theorems for Optimization} {No free lunch
  theorems for optimization}.{\BBCQ}
\newblock
\APACjournalVolNumPages{IEEE Transactions on Evolutionary
  Computation}{1}{1}{67--82}.
\newblock
\begin{APACrefDOI} \doi{10.1109/4235.585893} \end{APACrefDOI}
\PrintBackRefs{\CurrentBib}

\bibitem [\protect \citeauthoryear {%
Zunino%
, Ghirotto%
, Armadillo%
\BCBL {}\ \BBA {} Fichtner%
}{%
Zunino%
\ \protect \BOthers {.}}{%
{\protect \APACyear {2022}}%
}]{%
zuninoHamiltonianMonteCarlo2022}
\APACinsertmetastar {%
zuninoHamiltonianMonteCarlo2022}%
\begin{APACrefauthors}%
Zunino, A.%
, Ghirotto, A.%
, Armadillo, E.%
\BCBL {}\ \BBA {} Fichtner, A.%
\end{APACrefauthors}%
\unskip\
\newblock
\APACrefYearMonthDay{2022}{}{}.
\newblock
{\BBOQ}\APACrefatitle {Hamiltonian {{Monte Carlo Probabilistic Joint
  Inversion}} of {{2D}} (2.{{75D}}) {{Gravity}} and {{Magnetic Data}}}
  {Hamiltonian {{Monte Carlo Probabilistic Joint Inversion}} of {{2D}}
  (2.{{75D}}) {{Gravity}} and {{Magnetic Data}}}.{\BBCQ}
\newblock
\APACjournalVolNumPages{Geophysical Research Letters}{49}{20}{e2022GL099789}.
\newblock
\begin{APACrefDOI} \doi{10.1029/2022GL099789} \end{APACrefDOI}
\PrintBackRefs{\CurrentBib}

\bibitem [\protect \citeauthoryear {%
Zunino%
\ \BBA {} Mosegaard%
}{%
Zunino%
\ \BBA {} Mosegaard%
}{%
{\protect \APACyear {2018}}%
}]{%
zuninoIntegratingGradientInformation2018}
\APACinsertmetastar {%
zuninoIntegratingGradientInformation2018}%
\begin{APACrefauthors}%
Zunino, A.%
\BCBT {}\ \BBA {} Mosegaard, K.%
\end{APACrefauthors}%
\unskip\
\newblock
\APACrefYearMonthDay{2018}{{\APACmonth{06}}}{}.
\newblock
{\BBOQ}\APACrefatitle {Integrating {{Gradient Information}} with
  {{Probabilistic Traveltime Tomography Using}} the {{Hamiltonian Monte Carlo
  Algorithm}}} {Integrating {{Gradient Information}} with {{Probabilistic
  Traveltime Tomography Using}} the {{Hamiltonian Monte Carlo
  Algorithm}}}.{\BBCQ}
\newblock
\BIn{} \APACrefbtitle {80th {{EAGE Conference}} \& {{Exhibition}} 2018
  {{Workshop Programme}}} {80th {{EAGE Conference}} \& {{Exhibition}} 2018
  {{Workshop Programme}}}\ (\BPG~cp).
\newblock
\APACaddressPublisher{}{{European Association of Geoscientists \& Engineers}}.
\newblock
\begin{APACrefDOI} \doi{10.3997/2214-4609.201801971} \end{APACrefDOI}
\PrintBackRefs{\CurrentBib}

\bibitem [\protect \citeauthoryear {%
Zunino%
\ \BBA {} Mosegaard%
}{%
Zunino%
\ \BBA {} Mosegaard%
}{%
{\protect \APACyear {2019}}%
}]{%
zuninoEfficientMethodSolve2019}
\APACinsertmetastar {%
zuninoEfficientMethodSolve2019}%
\begin{APACrefauthors}%
Zunino, A.%
\BCBT {}\ \BBA {} Mosegaard, K.%
\end{APACrefauthors}%
\unskip\
\newblock
\APACrefYearMonthDay{2019}{{\APACmonth{01}}}{}.
\newblock
{\BBOQ}\APACrefatitle {An Efficient Method to Solve Large Linearizable Inverse
  Problems under {{Gaussian}} and Separability Assumptions} {An efficient
  method to solve large linearizable inverse problems under {{Gaussian}} and
  separability assumptions}.{\BBCQ}
\newblock
\APACjournalVolNumPages{Computers \& Geosciences}{122}{}{77--86}.
\newblock
\begin{APACrefDOI} \doi{10.1016/j.cageo.2018.09.005} \end{APACrefDOI}
\PrintBackRefs{\CurrentBib}

\bibitem [\protect \citeauthoryear {%
Zunino%
, Mosegaard%
, Lange%
, Melnikova%
\BCBL {}\ \BBA {} Mejer~Hansen%
}{%
Zunino%
\ \protect \BOthers {.}}{%
{\protect \APACyear {2015}}%
}]{%
zuninoMonteCarloReservoir2015}
\APACinsertmetastar {%
zuninoMonteCarloReservoir2015}%
\begin{APACrefauthors}%
Zunino, A.%
, Mosegaard, K.%
, Lange, K.%
, Melnikova, Y.%
\BCBL {}\ \BBA {} Mejer~Hansen, T.%
\end{APACrefauthors}%
\unskip\
\newblock
\APACrefYearMonthDay{2015}{{\APACmonth{01}}}{}.
\newblock
{\BBOQ}\APACrefatitle {Monte {{Carlo}} Reservoir Analysis Combining Seismic
  Reflection Data and Informed Priors} {Monte {{Carlo}} reservoir analysis
  combining seismic reflection data and informed priors}.{\BBCQ}
\newblock
\APACjournalVolNumPages{Geophysics}{80}{}{31}.
\newblock
\begin{APACrefDOI} \doi{10.1190/geo2014-0052.1} \end{APACrefDOI}
\PrintBackRefs{\CurrentBib}

\end{thebibliography}

\end{document}